\documentclass[12pt,a4paper,final]{iopart}
\usepackage{iopams, bm}

\makeatletter
\long\def\@makefntext#1{\parindent 1em\noindent 
 \makebox[1em][l]{\footnotesize\rm$\m@th{^\arabic{footnote}}$}%
 \footnotesize\rm #1}
\def\@makefnmark{\hbox{$^{\arabic{footnote}}\m@th$}}
\def\@thefnmark{\arabic{footnote}}
\makeatother
\usepackage{graphicx}

\usepackage{siunitx}
\usepackage{braket}

\usepackage{xcolor}

\newcommand{\gstate}{\mathrm{X}^{2}\Sigma^{+}}
\newcommand{\estate}{\mathrm{A}^{2}\Pi_{1/2}}

\begin{document}

\title{An ultracold molecular beam for testing fundamental physics}

\author{X. Alauze\footnote{These authors contributed equally to this work.}\addtocounter{footnote}{-1}, J. Lim\footnotemark, M. A. Trigatzis, S. Swarbrick, F. J. Collings, N. J. Fitch, B. E. Sauer and M. R. Tarbutt}

\address{Centre for Cold Matter, Blackett Laboratory, Imperial College London, Prince Consort Road, London SW7 2AZ UK}
\eads{x.alauze@imperial.ac.uk, j.lim@imperial.ac.uk, m.tarbutt@imperial.ac.uk}

\begin{abstract}
We use two-dimensional transverse laser cooling to produce an ultracold beam of YbF molecules. Through experiments and numerical simulations, we study how the cooling is influenced by the polarization configuration, laser intensity, laser detuning and applied magnetic field. The ultracold part of the beam contains more than $2 \times 10^5$ molecules per shot and has a temperature below 200~${\mu}$K, and the cooling yields a 300-fold increase in the brightness of the beam. The method can improve the precision of experiments that use molecules to test fundamental physics. In particular, the beam is suitable for measuring the electron electric dipole moment with a statistical precision better than $10^{-30}$~e~cm.
\end{abstract}

\noindent{\it Keywords}: laser cooling molecules, electron electric dipole moment

\section{Introduction}\label{sec:intro}

Despite its many successes, e.g.~\cite{Abachi1995,AbeF2014,Kodama2001,Aad2012,Chatrchyan2012}, the Standard Model of particle physics fails to account for several major cosmological observations, including dark matter, the accelerating expansion of the Universe (dark energy), and the asymmetry between the amount of matter and antimatter in the Universe \cite{Faber1979,Dine2003,Weinberg1989}. It is therefore widely accepted that there must be physics beyond the Standard Model. One of the requirements for matter-antimatter asymmetry is CP violation \cite{Sakharov1991b}. The Standard Model violates CP symmetry, but the mechanism cannot account for the observed baryon asymmetry \cite{Dine2003}. In all CPT-invariant theories, CP violation is equivalent to T-violation and so introduces T-violating phenomena. A fundamental particle with an electric dipole moment (EDM) is the most notable example of such T-violation, but no EDM has yet been measured for any particle.
In the Standard Model, the electron acquires an EDM via the CP-violating phase that appears in the Cabbibo–Kobayashi–Maskawa model of quark flavour mixing. This electron electric dipole moment (eEDM, $d_e$) is predicted to be less than $10^{-44}$~e~cm~\cite{Pospelov2014}, which is an extremely tiny value. A CP-violating electron-nucleon interaction can be interpreted as an effective $d_e$, since it has the same experimental signature, but this value is predicted to be less than $10^{-38}$~e~cm, which is still tiny. Theories that extend the Standard Model typically introduce new CP-violating interactions which could help explain the baryon asymmetry. These new interactions also tend to generate much larger eEDM values, often in the range of $10^{-30}$--$10^{-24}$~e~cm ~\cite{Khriplovich1997}. The present experimental upper limit on the eEDM is $|d_e|<1.1\times10^{-29}$ e cm \cite{Andreev2018}. This result sets tight constraints on the parameter spaces of many theories beyond the Standard Model up to an energy scale of about 30~TeV. The eEDM is thus a powerful probe of new physics, and the prospect of measuring the eEDM with even higher precision is an exciting one.

Over the last decade, all of the most precise eEDM measurements have been made using heavy polar molecules~\cite{Hudson2011,Kara2012,Baron2014,Cairncross2017,Andreev2018}. In these systems, the interaction energy is $U = -d_e \vec{\sigma}\cdot \vec{E}_{\rm eff}$ where $\vec{\sigma}$ is the electron spin and $\vec{E}_{\rm eff}$ is an effective electric field. For the molecules typically used, $E_{\rm eff}$ is in the range of 10--100~GV/cm when the molecule is fully polarized, and complete polarization requires only a modest laboratory electric field. So far, two experimental approaches have been successful, one using beams of neutral molecules~\cite{Hudson2011, Baron2014,Andreev2018} where the number of molecules can be large but the coherence time is limited to a few milliseconds by the flight time and the divergence of the beam, and another using trapped molecular ions where the coherence time can be three orders of magnitude larger but the number of ions is small~\cite{Cairncross2017, Zhou2020}. For neutral molecules, the coherence time could be greatly improved by using laser cooling to produce a highly collimated, slowly moving molecular beam, or an ensemble of trapped ultracold molecules~\cite{Tarbutt2013, Fitch2020b}. This is the approach we pursue here.

Because of their more complex energy level structure, laser cooling of molecules is more difficult than for atoms. Nevertheless, many of the techniques commonly used to produce and manipulate ultracold atoms -- Doppler cooling, sub-Doppler cooling, magneto-optical traps, magnetic traps, optical dipole traps, collisional cooling -- have all recently been applied to molecules. These advances are reviewed in references~\cite{Tarbutt2018, McCarron2018b, Fitch2021}. Some molecules, including YbF, BaF, and YbOH, are good species for eEDM measurements and also suitable for laser cooling. One dimensional transverse laser cooling has been demonstrated for YbF~\cite{Lim2018} and YbOH~\cite{Augenbraun2020}, showing great promise for a new generation of eEDM measurements based on ultracold molecules. The YbF molecule has been used to measure the eEDM in the past~\cite{Hudson2002, Hudson2011}, and recent work has shown how to distil the population into a single rotational state and how to detect these molecules with high efficiency and state selectivity~\cite{Ho2020}. In this paper, we demonstrate two dimensional laser cooling of YbF. We study two sub-Doppler mechanisms: magnetically induced laser cooling and polarization gradient cooling. Both mechanisms were understood more than 30 years ago for atoms~\cite{Sheehy1990,Emile1993,Dalibard1989,Lett1989,Grynberg1994} and have been used for one-dimensional transverse cooling of molecules~\cite{Shuman2010,Kozyryev2017,Augenbraun2020,Mitra2020}. As the molecules travel through intensity and/or polarization gradients produced by counter-propagating laser beams, they climb potential energy hills formed by the spatial modulation of the ac Stark shift. When the detuning is positive, they tend to be optically pumped to a dark state near the top of the hills, and returned to a bright state near the bottom of the hills, either by Larmor precession (magnetically induced cooling) or due to their motion through the changing polarization of the light field (polarization gradient cooling). The molecules are cooled because they repeatedly climb the potential hills. Here, we explore the parameter space of the laser cooling in detail, through both experiment and simulation, and find parameters that produce an intense, highly collimated molecular beam. We assess these results in the context of an eEDM measurement.

\section{Laser cooling scheme and setup of the experiment}\label{sec:setup}

\begin{figure}[tb]
\centering
\includegraphics[width=0.5\textwidth]{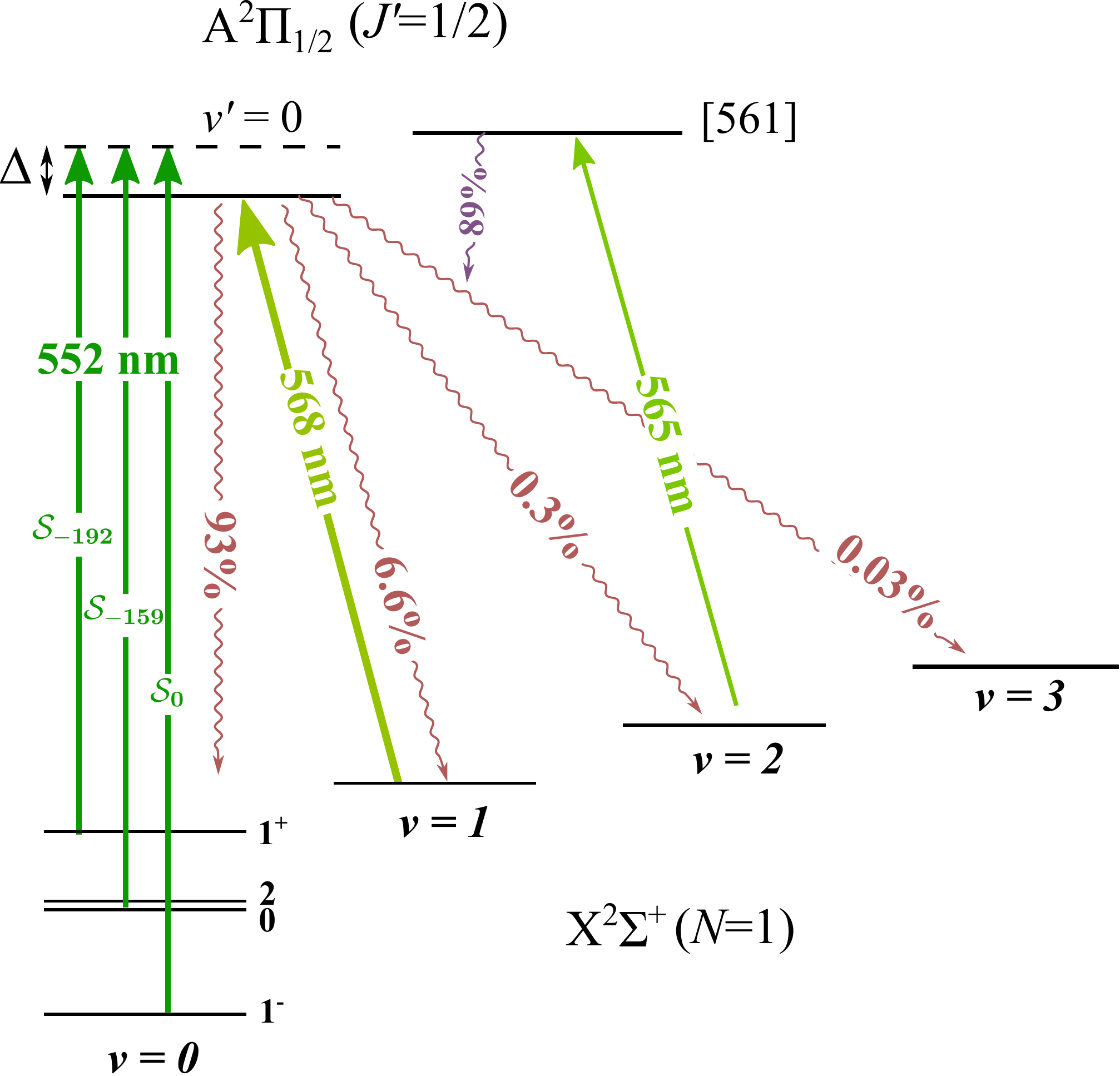}
\caption{YbF energy levels involved in the laser cooling scheme. The vibrational branching ratios are shown (wavy arrows) together with the cooling and repumping transitions (solid arrows). The three hyperfine components of the main cooling transition are shown, and are labelled ${\cal S}_{\delta}$ with $\delta$ the detuning (in MHz) relative to the sideband of highest frequency. The sidebands applied to the repump lasers are not shown. 

}
\label{fig:CoolingLevels}
\end{figure}

Figure \ref{fig:CoolingLevels} shows the energy levels of the YbF molecule involved in the laser cooling scheme. The electronic ground and first excited states are labelled as $\gstate$ and $\estate$, respectively. We introduce the following quantum numbers: $v$ labels the vibrational state, $N$ is the rotational angular momentum, $J$ is the total angular momentum of the molecule apart from nuclear spin, $I=1/2$ is the spin of the fluorine nucleus, $F$ is the total angular momentum, and $m_F$ is its projection onto the quantization axis. We use primes for quantum numbers in the excited state. 
The main cooling transition is $\ket{\estate,v'=0, J'=1/2} - \ket{\gstate,v=0,N=1}$ which has a wavelength of $\lambda = 552$~nm, a natural linewidth of $\Gamma = 2\pi \times 5.7$~MHz, and a saturation intensity of $I_{\rm sat} = \pi h c \Gamma/(3 \lambda^3) = 4.4$~mW/cm$^2$. From this excited state, electric dipole transitions to any other rotational states of $\gstate$ are forbidden by parity and angular momentum selection rules, but decay to other vibrational states are allowed. Their branching ratios~\cite{Zhuang2011,Smallman2014} are indicated in figure~\ref{fig:CoolingLevels}. Population that decays to $v=1$ and $v=2$ is returned to the cooling cycle by driving the $\ket{\estate,v'=0, J'=1/2} - \ket{\gstate,v=1,N=1}$ transition at 568~nm and the $\ket{[561], J'=1/2} - \ket{\gstate,v=2,N=1}$ transition at 565~nm. The state [561]\footnote{These mixed states are labelled using the notation $[E]$ where $E$ is the approximate energy of the state, in THz, above the ground state.} is a mixture of the $\estate(v'=1)$ state and a different electronic state arising from inner-shell excitation of the molecule~\cite{Lim2017}. 

The laser system used to implement this cooling scheme is described in detail in \cite{Trigatzis2020}. We use the notation ${\cal L}_v$ to refer to the lasers where the subscript $v$ represents the vibrational quantum number in the ground electronic state that the molecule addresses. In this notation, the main cooling laser at 552~nm is ${\cal L}_0$ while the two vibrational repump lasers are ${\cal L}_1$ and ${\cal L}_2$. Since our previous 1D cooling work~\cite{Lim2018} we have upgraded ${\cal L}_0$ to provide substantially more power. The power incident on the molecules from each of these lasers is $P_0$, $P_1$ and $P_2$ with maximum values of 850~mW, 300~mW and 15~mW respectively. 
A laser designed to close the leak to $v=3$ is also available but made no significant difference to laser cooling so was not used for any of the results presented here. Each vibrational state of $\gstate$ shown in figure \ref{fig:CoolingLevels} has four components arising from the spin-rotation and hyperfine interactions.\footnote{Even though there are two different interactions involved, we refer to these simply as `hyperfine components'. They are only shown for $v=0$ in figure~\ref{fig:CoolingLevels}.}  These have $F=0, 1, 1$ and 2. We distinguish the two $F=1$ levels using the notation $1^{\pm}$, where $1^{+}$ has higher energy. The hyperfine components are addressed by adding rf sidebands to each laser using a combination of acousto-optic modulators (AOMs) and electro-optic modulators (EOMs). In the case of ${\cal L}_0$, we use only three rf sidebands as the spacing between the $F=0$ and $F=2$ levels is smaller than the natural linewidth of the transition. These three sidebands are labelled as ${\cal S}_{\delta}$ and have powers $r_{\delta} P_0$, where $\delta$ is the detuning in MHz relative to the sideband of highest frequency.

\begin{figure}[tb]
\centering
\includegraphics[width=0.9\textwidth]{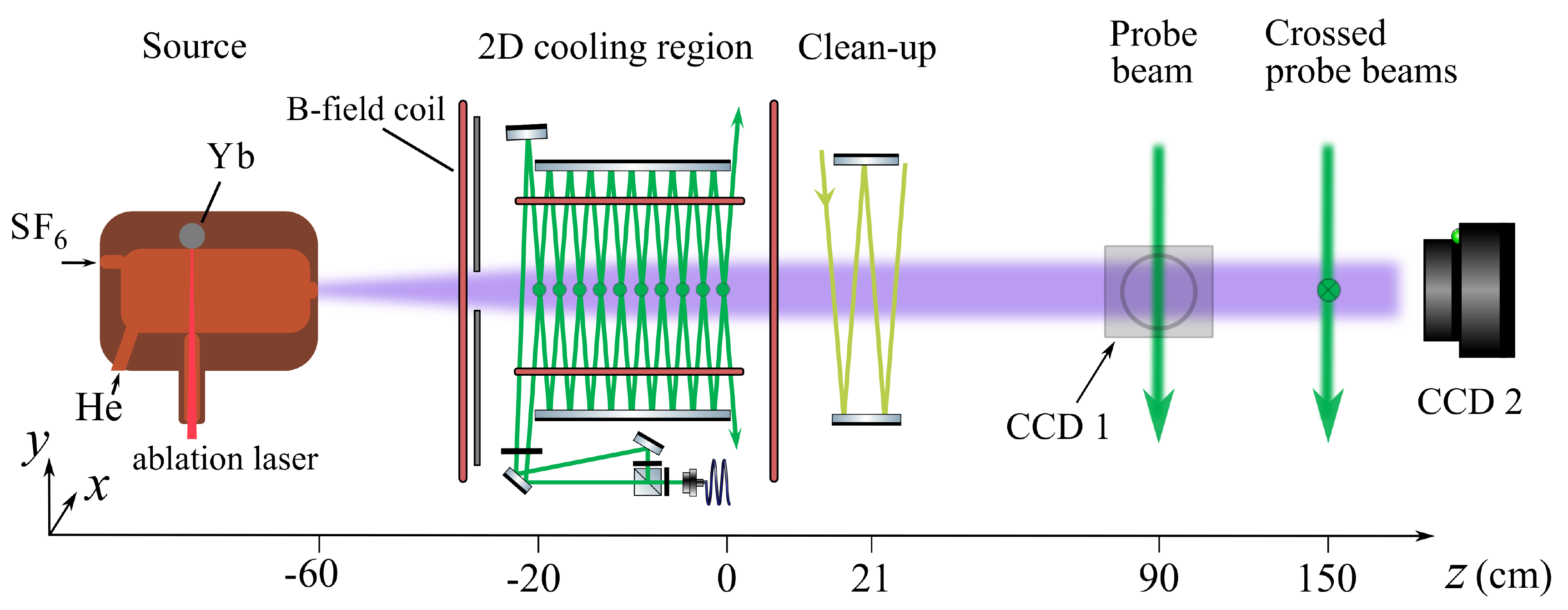}
\caption{Side sectional view of the experimental setup. The filled green circles in the cooling and probe regions represent the beams propagating in the horizontal $x$ direction. }
\label{fig:setup}
\end{figure}

Figure \ref{fig:setup} illustrates the setup of the experiment. The light from all three lasers, including rf sidebands, is combined into a single beam which is then divided into two beams of equal power. These beams are coupled into two single-mode fibres for delivery to the cooling region, where they are used for cooling in the two orthogonal transverse directions ($x$ and $y$). The figure shows the configuration of the cooling light in the $y$-direction. The beam from the fibre is split into two parts which then enter the cooling region from opposite sides with their polarizations parallel. The beams have a waist of 2.5~mm and bounce back and forth between a pair of mirrors thirty times to create a 20~cm long sheet of light. An identical arrangement is used in the $x$-direction, making a two-dimensional optical lattice. In the cooling region, the peak intensity at 552~nm (${\cal L}_0$) due to a single laser beam is denoted $I_{\rm max}$. This definition should be noted carefully, since there are other intensities that are relevant. For example, the intensity on the molecular beam axis ($x=y=0$), averaged over the length of the cooling region in the $z$-direction is considerably smaller, about $0.24 I_{\rm max}$. Conversely, since there are four beams and they may interfere, the peak intensity in the optical molasses is considerably higher.

We have studied two polarization configurations in detail. In the first, which we call $XY_{||}$, all four beams are polarized along $z$ so that the electric field is $(E_x, E_y, E_z)=(0,0, 2E_0 (\cos{kx}+\cos{ky}) \cos{\omega t})$. In the second, called $XY_{\perp}$, the beams propagating along $\pm x$ are polarized along $y$, while those propagating along $\pm y$ are polarized along $z$, so that the field is $(0, 2E_0 \cos{kx} \cos{\omega t}, 2E_0 \cos{ky} \cos{\omega t})$. In both polarization configurations, shim coils are used to cancel the background magnetic field in the cooling region and apply a uniform magnetic field, $B(0,\frac{1}{\sqrt{2}},\frac{1}{\sqrt{2}})$.

The beam of YbF molecules is produced using a cryogenic buffer gas source, which has been described in \cite{Lim2018} and is similar to the design of \cite{Truppe2017c}. Helium buffer gas flows through a copper cell cooled to 5~K. Yb atoms are introduced into the cell by laser ablation of a metal target using a pulsed Nd:YAG laser. The Yb reacts with SF$_6$ gas that is injected into the cell through a capillary held near room temperature, creating YbF molecules which then thermalize with the buffer gas. The source produces a molecular beam with a flux between $3 \times 10^{9}$ and $9 \times 10^{9}$ molecules per steradian per pulse in the fist rotational state ($N=1$), with a mean forward velocity between 160 and 220 m/s, depending on the He gas flow and ablation laser power. Defining $z=0$ to be at the exit of the cooling region, the molecules leave the cell at $z=-60$~cm, enter the cooling region at $z=-20$~cm, pass through the `clean-up' region at $z=21$~cm, and are detected at $z=90$ and 150~cm. In the clean-up region, which contains only repump light, all the population is pumped into $v=0$. In the two detectors, the spatial distribution of the molecules is measured by collecting laser-induced fluorescence onto an EMCCD (electron multiplying charge-coupled device). In the first region, only the distribution along $y$ is measured, while in the second we use two crossed probe beams to measure the distributions in both the $x$ and $y$ dimensions. An additional detector, not shown in the figure, can be introduced at $z=-20$~cm. We use photomultiplier tubes (PMTs) at $z_{\rm PMT1}=-20$~cm and  $z_{\rm PMT2}=90$~cm for time-resolved fluorescence detection which we use to determine the velocity distribution of the beam in the $z$ direction.

\section{Simulation results}
\label{sec:simulations}
In this section, we present the results of laser cooling simulations designed to match the experimental configurations. These simulations provide insight into the cooling and help interpret our experimental results that follow. We use 3D optical Bloch equation simulations~\cite{Devlin2016, Devlin2018} that include all the molecular levels of the $\ket{\estate, v'=0, J'=1/2)} - \ket{\gstate, v=0, N=1)}$ transition, and all frequency components of the 552~nm light. The leak to the higher-lying vibrational levels is neglected. The simulations predict how the parallel component of acceleration\footnote{Here, the parallel component refers to the component in the direction of the velocity.}, $a(v)$, and the excited state population, $n_{\rm e}(v)$, depend on the transverse velocity $v$. These are key results that can be used to determine the capture velocity of the optical molasses and the equilibrium temperature, so we study how they are influenced by the polarization configuration ($XY_{||}$ and $XY_{\perp}$), the detuning, $\Delta$, the laser intensity per beam, $I_0$, and the magnetic field $B$. For all simulations, we fix the relative intensities of the sidebands to $\{r_0, r_{-159}, r_{-192}\}=\{1/2,1/2,0\}$, which is the arrangement used for most of the experiments. Note that the intensity is a constant in the simulations, whereas in the experiments the intensity varies with position.

\subsection{Magnetic field and polarization} \label{sec:sim_Bfield_Polar}
\begin{figure}[tb]
\centering
\includegraphics[width=0.8\textwidth]{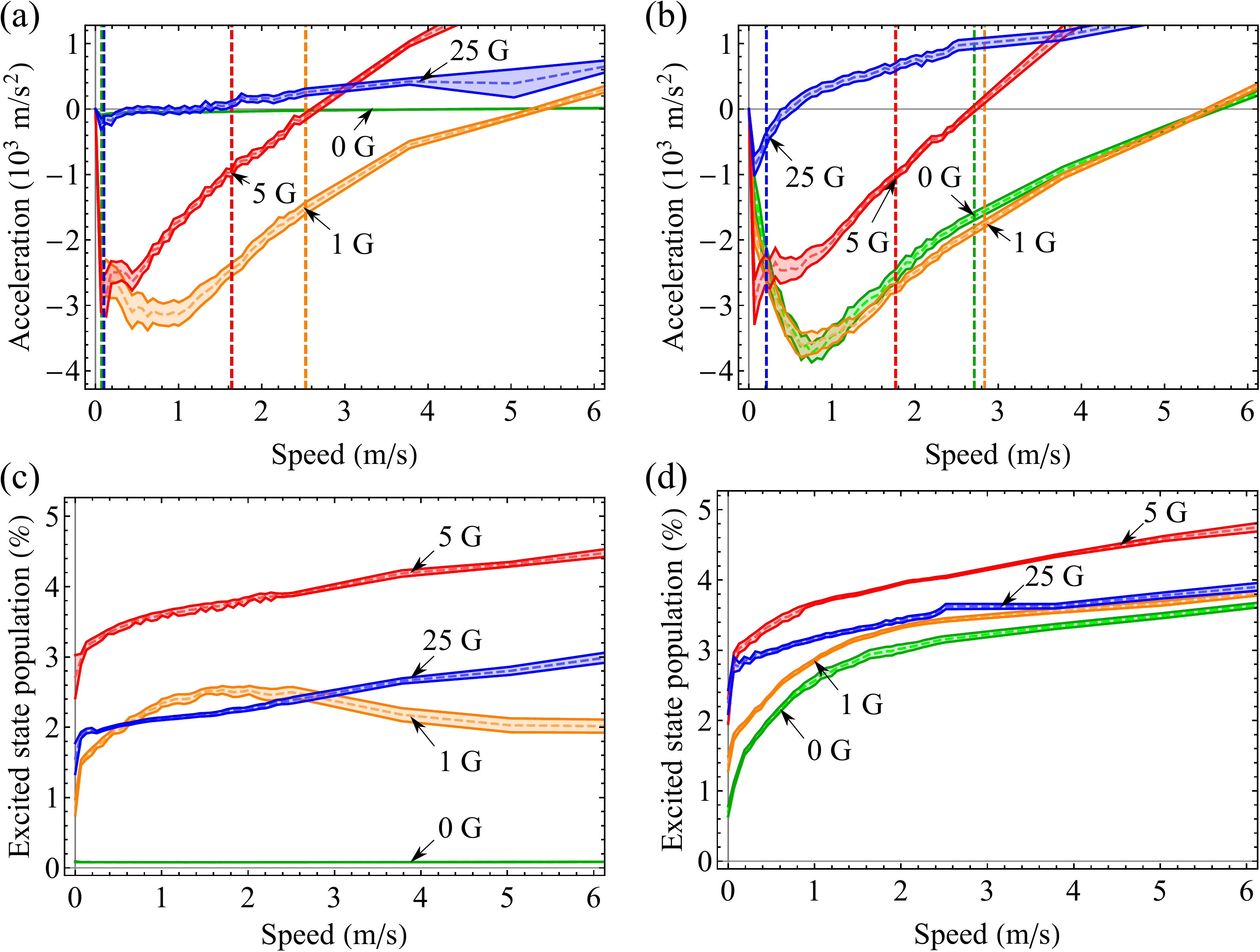}
\caption{(a,b) Parallel component of acceleration versus velocity for (a) $XY_{||}$ and (b) $XY_{\perp}$. (c,d) Excited state population as a function of speed for (c) $XY_{||}$ and (d) $XY_{\perp}$. The lines are for different values of $B$: 0~G (green), 1~G (orange), 5~G (red), and 25~G (blue). The intensity is $I_0 = 100\,I_{\rm sat}$, and the detuning is $\Delta = 6\, \Gamma$. For each curve, the shaded band indicates the 1-$\sigma$ confidence interval determined from the results of multiple simulations. Vertical dashed lines show the value of the critical velocity, $v_0$, for each curve. }
\label{fig:coolingForceMagneticField}
\end{figure}
Figure~\ref{fig:coolingForceMagneticField} shows how the parallel component of acceleration and the excited state population depend on the velocity for the two polarization configurations and for various values of $B$. The detuning is $\Delta = 6\Gamma$, and the intensity is $I_0=100I_{\rm sat}$,  which is characteristic of the intensities used in the experiments. In all cases, we see that the acceleration is positive at high speeds where Doppler heating dominates, and negative at lower speeds where sub-Doppler cooling forces dominate. This is a general feature of systems where $F \ge F'$. We define two characteristic velocities for the molasses. The first is the capture velocity, $v_{\rm capture}$, which is the velocity where the force crosses zero. The second is a critical velocity, $v_0$. This is the maximum velocity that a molecule can have for the velocity to be reduced to zero within $t_0 = 1$~ms, which is the typical interaction time in the experiment. It satisfies $\int_{v_0}^{0} dv/a(v) = t_0$ and is indicated by the vertical dashed lines in figure~\ref{fig:coolingForceMagneticField}(a,b).

For the $XY_{||}$ configuration, shown in  figure~\ref{fig:coolingForceMagneticField}(a) and (c), the excited state population and the acceleration are zero when $B=0$. This is because the polarization is uniform (parallel to $z$) throughout the cooling region, and the $m_F=\pm 2$ ground-state sublevels are dark states for this polarization. Molecules are optically pumped into these dark states where they remain indefinitely. When $B \ne 0$ the Larmor precession rotates the dark states into bright states and molecules can scatter photons continuously. Note, however, that this mechanism is suppressed when the frequency difference between the dark and bright states, due to the ac Stark shift, is larger than the Larmor frequency, so the dark states are destabilized most effectively near the nodes of the light field. This limits the scattering rate at low velocity, because the molecules take too long to reach the nodes. This is the reason for the rapid drop in the excited state population at very low velocities.  As $B$ increases from 0 to 5~G, the excited state population increases because the dark states are destabilized more rapidly. For fields of 1 and 5~G the damping force increases very rapidly as the velocity increases from 0 to 0.1~m/s, and then gradually decreases over a much wider range of velocities. The capture velocity, critical velocity and maximum cooling force are all greatest at $B=1$~G. To help understand these results, we introduce three characteristic time scales. The first is the optical pumping time, $\tau_{\rm  OP}$, which is the time taken to be pumped to the dark state. At the high intensities used here, this is limited by the decay rate of the excited state, so we take $\tau_{\rm OP} = 6/\Gamma$. The factor of 6 is the number of ground-state sublevels divided by the number of dark states. The second time scale is the time taken to travel from a node to an antinode of the standing wave, $\tau_{\rm travel} = \lambda/(4v)$. The third is the Larmor precession time $\tau_{\rm  Larmor} = \hbar/(g \mu_{\rm B} B)$. Here, $g=1/3$ is the degeneracy-weighted average $g$-factor of the ground-state hyperfine components. We expect the cooling to be most effective when $\tau_{\rm OP} \approx \tau_{\rm travel} \approx \tau_{\rm Larmor}$. This argument suggests that the cooling will be maximized when $v \approx 1$~m/s and $B \approx 2$~G. Reassuringly, the simulations are consistent with these rough estimates.

For the $XY_{\perp}$ configuration, shown in  figure~\ref{fig:coolingForceMagneticField}(b) and (d), the polarization is not uniform but varies periodically in the $x$ and $y$ directions. The motion of the molecules through this spatially varying polarization results in nonadiabatic transitions between dark and bright states. Photon scattering continues without the aid of a magnetic field, as can be seen from the excited population which does not go to zero at $B=0$. The excited population tends to zero as both $v$ and $B$ go to zero, for then there ceases to be a way out of the dark states. This reduction at small $v$ also happens when $B \ne 0 $, but the population does not go to zero in this case. For small $B$, the cooling force is maximized at around the same velocity as for the $XY_{||}$ case. The acceleration curve hardly changes from 0 to 1~G, but increasing $B$ beyond 1~G lowers the maximum acceleration, the capture velocity and the critical velocity, so we expect magnetic fields larger than this to be detrimental to cooling in this configuration. 

We note that, despite the complex level structure of the molecule, the results presented in figure~\ref{fig:coolingForceMagneticField} show similar features to those found for simpler systems where $F \ge F'$~\cite{Devlin2016}. This suggests that the cooling in this case can indeed be understood in the context of the usual simple models of sub-Doppler cooling. We also find that the $XY_{||}$ configuration shows similar behaviour to lin$\parallel$lin cooling in 1D, and that the $XY_{\perp}$ case is similar to lin$\perp$lin cooling in 1D~\cite{Lim2018}.

\subsection{Detuning and laser intensity}
\begin{figure}[tb]
\centering
\includegraphics[width=0.8\textwidth]{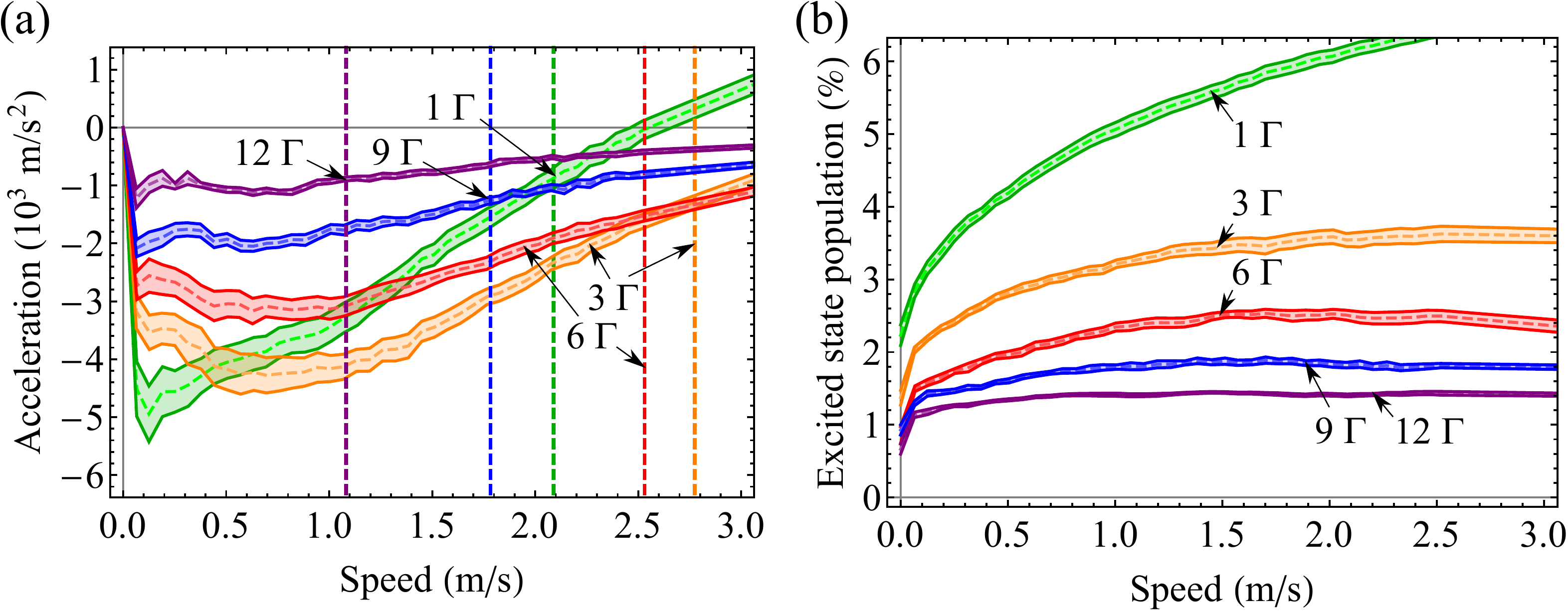}
\caption{(a) Parallel component of acceleration and (b) excited state population as a function of velocity for three values of detuning, $\Delta$:  $1\Gamma$ (green), $3 \Gamma$ (orange), $6\Gamma$ (red), $9\Gamma$ (blue), and $12\Gamma$ (purple). The polarization configuration is $XY_{||}$ and $B = 1$~G. The intensity is $I_0=100 I_{\rm sat}$. Vertical dashed lines show the value of the critical velocity, $v_0$, for each curve. }
\label{fig:coolingForceDetuning}
\end{figure}

Figure~\ref{fig:coolingForceDetuning} presents simulation results for various values of detuning. Here, the intensity is $I_0 =100 I_{\rm sat}$, the magnetic field is $B=1$~G and the polarization configuration is $XY_{||}$. While $v_{\rm capture}$ increases with $\Delta$ all the way to $12 \Gamma$, $v_0$ is largest for $\Delta = 3\Gamma$, suggesting that a detuning near this value will capture the most molecules, at least for this intensity. Repeating these simulations for other intensities shows that the optimum detuning increases as the intensity increases. For example, at $I_0 =300 I_{\rm sat}$ the optimum detuning is close to $6 \Gamma$. It is remarkable that the cooling force extends to very high velocities at large detuning and intensity. For example, when $\Delta = 12 \Gamma$, $v_{\rm capture} \approx 6$~m/s at $I_0 = 100 I_{\rm sat}$, increasing to 12~m/s at $300 I_{\rm sat}$. This suggests that a large fraction of the molecular beam could be captured and cooled to low temperature by using high intensity and large detuning. However, to make effective use of the large capture velocity, the interaction time with the laser light must be long enough. Using the acceleration curve shown in figure~\ref{fig:coolingForceDetuning}(a) for $\Delta = 12\Gamma$, we estimate that it will take 35~ms to reduce the speed from 6~m/s to below 1~m/s. The excited state population, shown in figure~\ref{fig:coolingForceDetuning}(b) drops at very low velocity for the same reason as discussed in the context of figure~\ref{fig:coolingForceMagneticField}(c). At higher velocities it tends towards a constant value because the Larmor precession time is the limiting factor. As we would expect, the excited state population decreases as the detuning increases. We note that simulations at other values of $B$ show similar features to those presented here.

\begin{figure}[tb]
\centering
\includegraphics[width=0.8\textwidth]{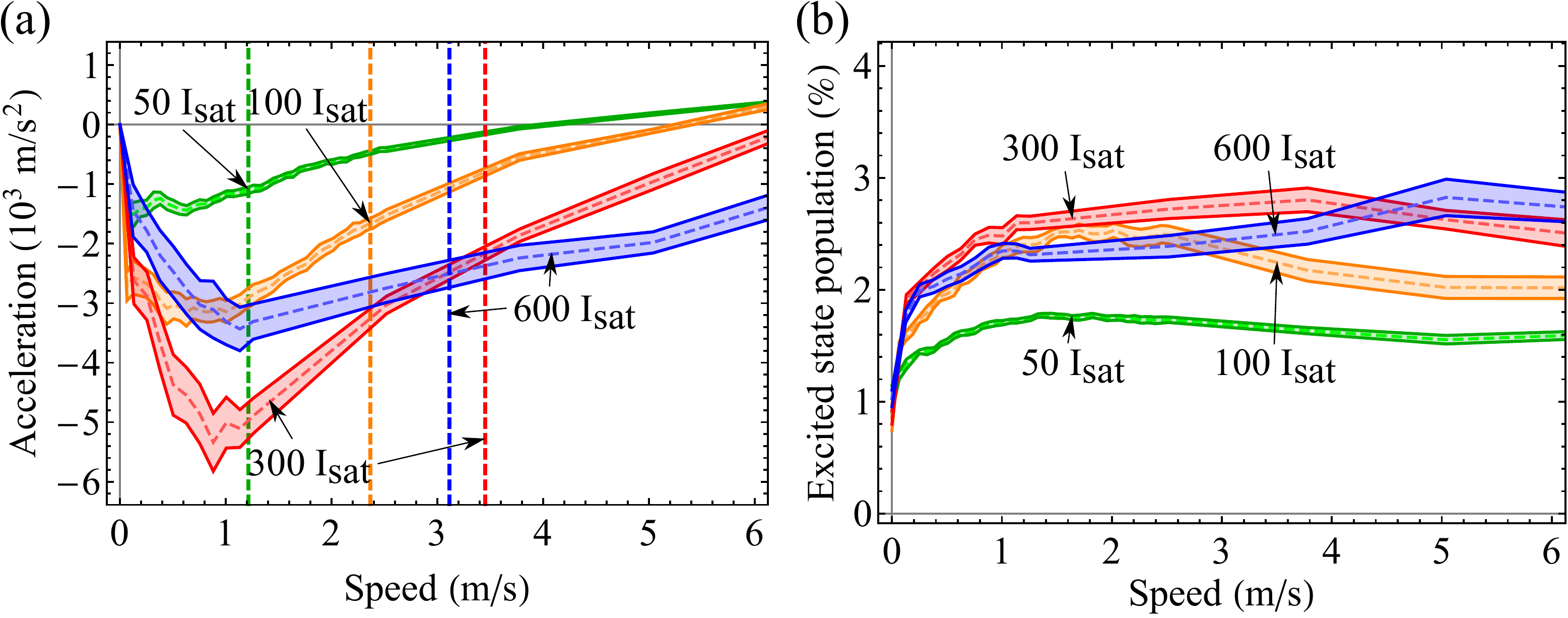}
\caption{Parallel component of (a) acceleration and (b) excited state population as a function of velocity for four values of intensity, $I_0$:  $50 I_{\rm sat}$ (green), $100 I_{\rm sat}$ (orange), $300 I_{\rm sat}$ (red), and $600 I_{\rm sat}$ (blue). The polarization configuration is $XY_{||}$, $B = 1$~G, and the detuning is $\Delta = 6 \Gamma$. Vertical dashed lines show the value of the critical velocity, $v_0$, for each curve. }
\label{fig:coolingForceIntensity}
\end{figure}

Figure \ref{fig:coolingForceIntensity} shows how the acceleration and excited state population depend on velocity for several values of intensity. Here, the polarization configuration is $XY_{||}$, $B = 1$~G, and $\Delta = 6 \Gamma$. We see that $v_{\rm capture}$ increases continuously with intensity. However, the maximum acceleration and the critical velocity, $v_0$, both increase up to $I_0 = 300 I_{\rm sat}$ and then decrease again. This is an important result, showing that when the cooling time is limited the number of ultracold molecules obtained does not increase indefinitely with laser intensity. The excited state population increases between 50 and $100 I_{\rm sat}$, but does not increase beyond $100 I_{\rm sat}$ because the limiting factor is the destabilization of dark states by the magnetic field.

Most of the experiments described in this paper use the $XY_{||}$ configuration and have a cooling time limited to about 1~ms. For this case, our simulations suggest that we will obtain the most ultracold molecules when the intensity is around $300 I_{\rm sat}$, the detuning is near $6\Gamma$ and the magnetic field is a few gauss. 

\subsection{Temperature evolution} \label{sec:sim_temperatureEvolution}
\begin{figure}[tb]
\centering
\includegraphics[width=0.8\textwidth]{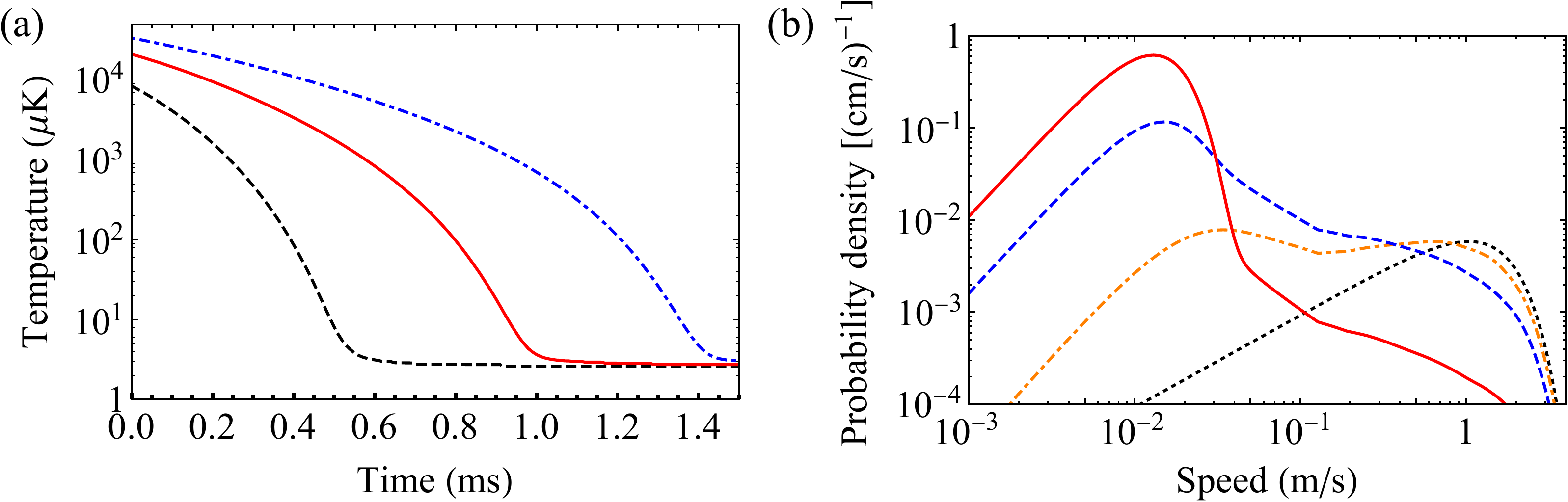}
\caption{(a) Transverse temperature as a function of cooling time for three initial temperatures: (black dashed) 10 mK, (red solid) 25 mK, and (blue dot-dashed) 40 mK. (b) Transverse speed distribution at four different cooling times: 0 (black dotted), 100 $\mu$s (orange dot-dashed), 300 $\mu$s (blue dashed), and 1 ms (red solid). The polarization configuration is $XY_{||}$, $\Delta = 6\Gamma$, $I_0=100 I_{\rm sat}$, and $B = 1$~G. The initial temperature is 25 mK.}
\label{fig:speedDistribution}
\end{figure}

Using the acceleration and excited state population results, the velocity distribution can be calculated by solving the Fokker-Planck-Kramers equation. Given the distance from source to laser cooling region, the cooling laser beam diameter, and the forward speed distribution of the molecules, we estimate that molecules in the cooling region have an initial transverse speed distribution that is close to a Maxwell-Boltzmann distribution at a temperature of 25~mK. The lower and upper bounds of this estimate are 10 and 40~mK, corresponding to velocity distributions with widths (FWHMs) of 1.5 and 3.1~m/s respectively. This range arises from the range of forward velocities and some uncertainty about the effective transverse extent of the laser cooling region. Figure~\ref{fig:speedDistribution}(a) shows the temperature as a function of cooling time for these three initial temperatures and for the parameters given in the caption. Because the tail of the speed distribution can skew the average speed enormously, we use a 5\% trimmed mean to determine a temperature. In the experiments, where the cooling time is about 1~ms, the model predicts that molecules will cool to about 3~$\mu$K when the initial temperature is 25~mK. Extending the cooling time to 1.4~ms brings the temperature below 10~$\mu$K even when the initial temperature is as high as 40~mK. As shown in Figures~\ref{fig:coolingForceMagneticField}(c) and \ref{fig:coolingForceDetuning}(b), the excited state population drops rapidly at the lowest speeds due to velocity selective coherent population trapping~\cite{Aspect1988,Caldwell2019}. The low photon scattering rate in this regime results in a low heating rate, and thus a low equilibrium temperature. Figure~\ref{fig:speedDistribution}(b) shows the transverse speed distribution at four different times, starting from a thermal distribution at 25~mK. A substantial fraction of the molecules reach a speed below 0.1~m/s within 300~$\mu$s, but there is a still a long tail that extends to high speed. After 1~ms, almost all the molecules have speeds below 0.1~m/s.

\subsection{Trajectory simulations}
\label{sec:traj_sim}

\begin{figure}[tb]
\centering
\includegraphics[width=0.8\textwidth]{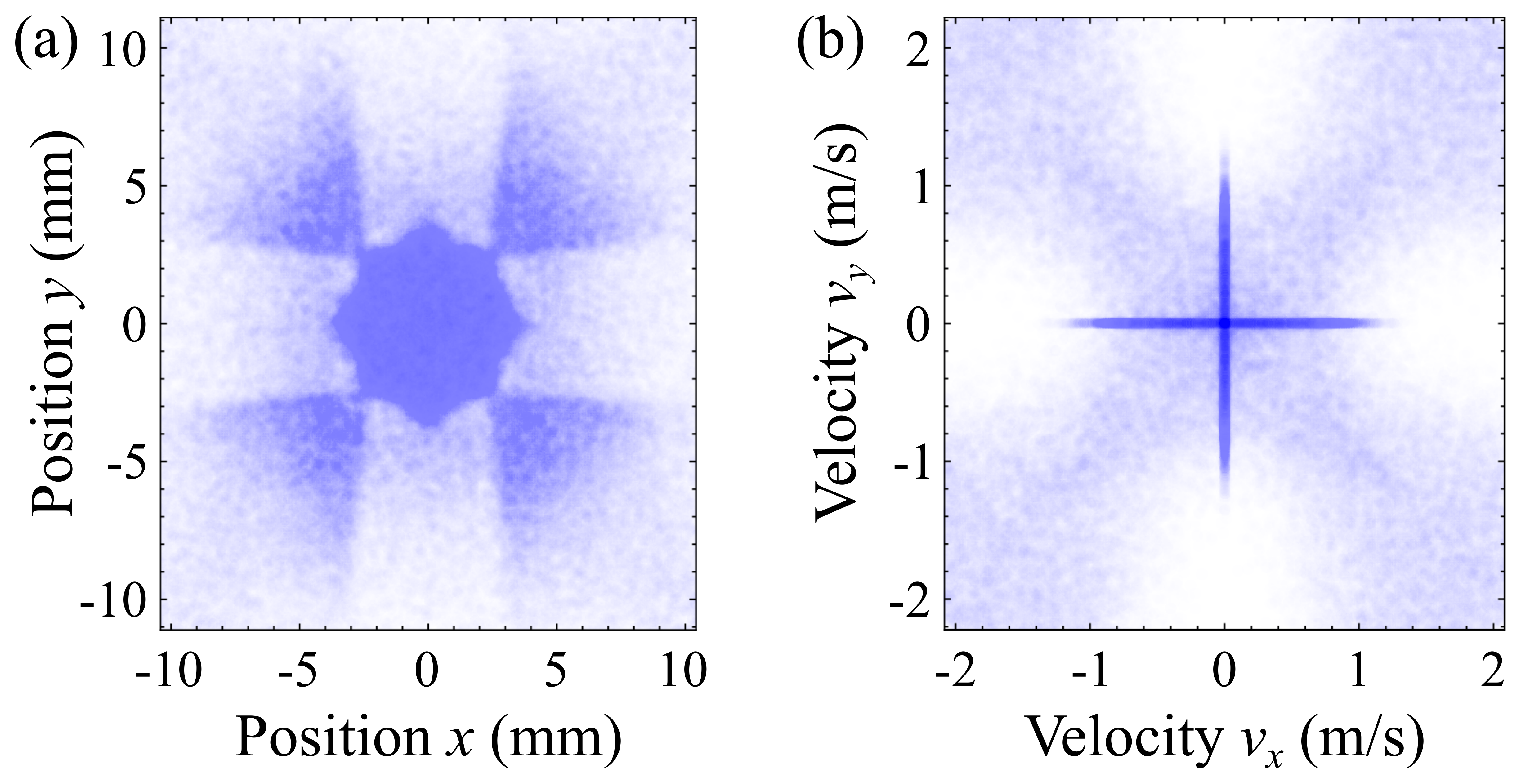}
\caption{Monte-Carlo trajectory simulations of 2D transverse laser cooling for the setup shown in figure~\ref{fig:setup}. (a) Transverse spatial distribution. (b) Transverse velocity distribution.}
\label{fig:trajectorySimulation}
\end{figure}

For comparison with the experimental results presented later, we perform Monte-Carlo trajectory simulations of the setup shown in figure~\ref{fig:setup} using the acceleration curves obtained from the optical Bloch equation simulations presented above. The molecules are generated from the buffer gas source with a Gaussian velocity distribution in the $z$-direction that has a mean of 200~m/s and a full width at half maximum (FWHM) of 60~m/s. In the transverse direction, we choose a Maxwell-Boltzmann velocity distribution with a temperature of 4~K. The initial spatial distribution is a 2D Gaussian with a FWHM of 2.5~mm cut off at a radius of 1.5~mm. The molecules travel freely for 40~cm, then enter the 20~cm long laser cooling region where they pass through 30 laser cooling beams with a Gaussian beam waist of 2.5~mm. After the laser cooling region, the molecules fly freely for a further 150~cm. The trajectories in the cooling region are determined solely from the calculated acceleration; the random momentum kicks are not included. We choose the polarization configuration to be $XY_{||}$, $B=1$~G and $\Delta = 6 \Gamma$.  Using the optical Bloch equations we calculate a set of acceleration curves similar to those in figure~\ref{fig:coolingForceIntensity}(a), but over a wider range of intensities, and then interpolate to determine the acceleration as a function of both speed and intensity, $\vec{a}^{\rm OBE} = a^{\rm OBE}(v,I)\hat{v}$. The simulations are done for equal intensity from all four beams. However, the experiment is more complicated than this because the intensity from the beams propagating along $x$ is, in general, different to the intensity from the beams along $y$, depending on the coordinate $\{x,y\}$. It is beyond the scope of this paper to calculate the acceleration for all values of $\{x,y,z,v_x,v_y\}$ so an approximation is needed to bridge this gap.  The acceleration used in the simulation is $a^{\rm sim}_x = a^{\rm OBE}(v_x, I_x(y,z))$ and an equivalent expression for the $y$ direction. Here, $I_x$ is the intensity due to the beams propagating along $x$ and is $I_x = \sum_{n=1}^{30} I_0 e^{-2y^2/w^2}e^{-2(z-z_{\rm s}-n l)^2/w^2}$ where $z_s$ is the start of the cooling region, $w=2.5$~mm and $l=6.5$~mm.

Figure~\ref{fig:trajectorySimulation} shows the simulated spatial and velocity distributions at $z$=150~cm. In figure~\ref{fig:trajectorySimulation}(a) we see a dense distribution at the centre corresponding to a highly collimated beam with a diameter of about 6~mm. More than 98\% of the molecules in this circle are transversely cooled to ultracold temperatures. Further out, there are regions where the density is depleted that reflect the intensity distribution of the cooling beams. The missing molecules from these regions are the ones brought into the central circle. Figure~\ref{fig:trajectorySimulation}(b) shows the transverse velocity distribution. It contains a tiny, dense spot at the centre (too tiny to see in the figure) which contains about 32\% of all the molecules. The dense cross which {\it is} visible in the figure corresponds to molecules that are cooled only in one direction. These molecules are ones that are far enough from the origin that they miss one of the two pairs of cooling beams. Further out we see areas of velocity space that are almost empty. The molecules that originally occupied these areas have all been cooled.

\section{Experimental results}
\label{sec:results}

\subsection{1D cooling}\label{sec:1D}

\begin{figure}[tb]
\centering
\includegraphics[width=1\textwidth]{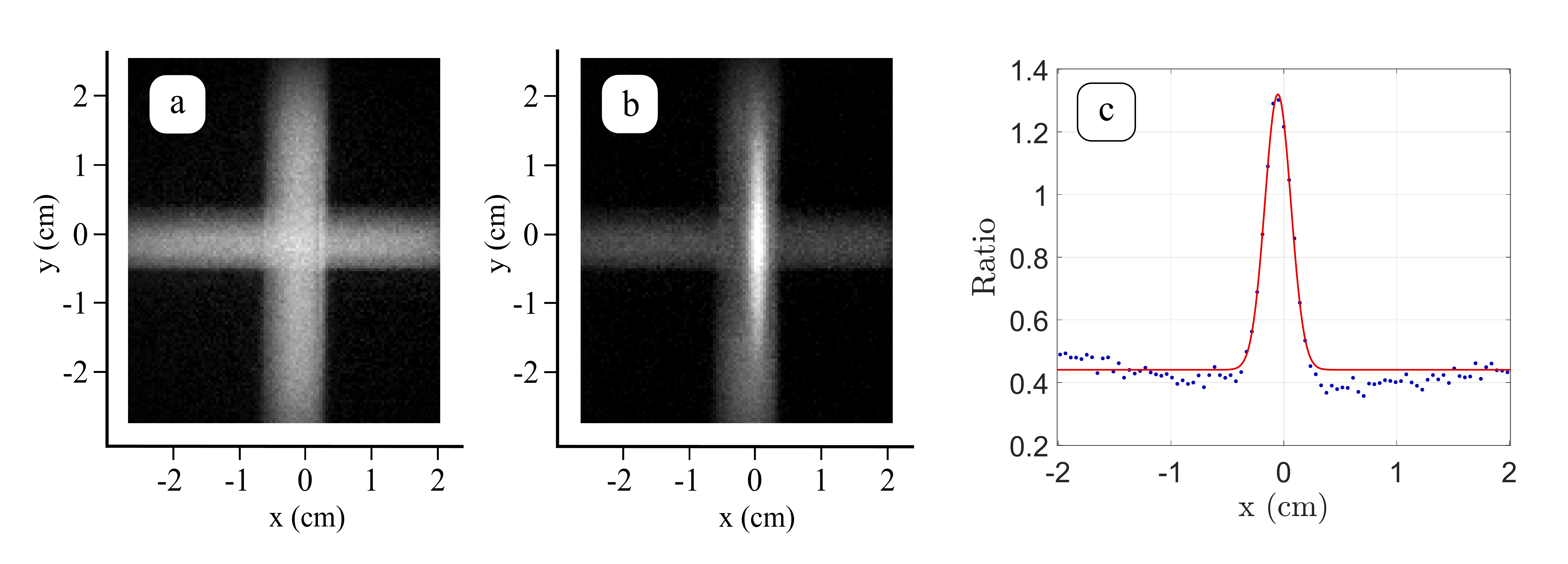}
\caption{Fluorescence images of molecules at CCD2, formed using two crossed probe beams. We use the polarization configuration $XY_{||}$, sideband configuration 5 of table \ref{tab:1Dsidebands} with a detuning of  $3\Gamma$, and a magnetic field of 0.5 G. The peak intensity per beam is $I_{\rm max} = 450 I_{\rm sat}$. (a) Distribution without laser cooling. (b) Distribution with laser cooling in 1D. (c) Ratio of the two images integrated along the $y$ direction. The line is a Gaussian fit to the data.}
\label{fig:1Dprofile}
\end{figure}

We start by cooling the molecules in each transverse direction separately, before moving on to 2D cooling. Figure~\ref{fig:1Dprofile} shows typical data for 1D cooling in the $x$ direction. We measure the density distribution by imaging laser-induced fluorescence onto CCD2 (see figure~\ref{fig:setup}). In these images, and in all our data, we subtract the background and average the results over 250 to 500 molecular beam pulses. Figure~\ref{fig:1Dprofile}(a) shows the image without laser cooling. At this distance from the source the density of molecules is uniform over the field of view of the camera, so the image reflects the intensity distribution of the two crossed probe beams. Figure \ref{fig:1Dprofile}(b) is the image of the cooled molecular beam. A bright vertical stripe appears, corresponding to molecules that have been cooled in the $x$ direction.  Figure~\ref{fig:1Dprofile}(c) shows the ratio of images (a) and (b), integrated along $y$. The points show the data and the line is a fit to the Gaussian model $g(x) = \frac{A}{\sqrt{2\pi}} e^{-(x-x_0)^2/(2\sigma^2)}+b$. We see a narrow peak whose area is proportional to the number of ultracold molecules and whose width is governed by both the critical velocity in the optical molasses and the final temperature obtained. We also notice that the baseline $b$ is well below 1. This loss of molecules is present for negative, positive and zero detuning, with $b$ typically lying between 0.4 and 0.7 depending on experimental parameters. This loss suggests that molecules are being optically pumped to a state that is not addressed by the laser light, i.e. there is a leak out of the cooling cycle. We are currently investigating the source of this leak -- it may be that population is decaying to an intermediate electronic state arising from inner-shell excitation, or to higher-lying vibrational levels ($v>3$) of $\gstate$. 

The area of the peak in figure \ref{fig:1Dprofile}(c), $A\sigma$, is proportional to the number of ultracold molecules at the detector. To make this dimensionless, we divide by a length $L$, giving us a useful figure of merit, $S_{\rm 1D} = A \sigma /L$. We only use $S_{\rm 1D}$ to evaluate the relative number of ultracold molecules obtained as a function of various parameters in the experiment, so the exact value of $L$ is not important. However, choosing $L$ to be the characteristic size of a molecular beam detector gives the following meaning to the absolute value of $S_{\rm 1D}$: when $S_{\rm 1D}=1$, the number of ultracold molecules at this detector is the same as the number with no laser cooling applied. We choose $L=5$~mm, which is the characteristic size of a laser-induced fluorescence detector optimized for high efficiency detection of a molecular beam. Later, we evaluate the brightness of the beam, which is a quantity that is independent of the distance from the source and combines all the desired qualities of a beam - small size, low divergence, and a large number of molecules.

\begin{table}[b]
\centering
\begin{tabular}{|c|c|c|c|c|c|c|}
\hline
Configuration   & $r_0$     & $r_{-159}$ & $r_{-192}$ & $\Delta$    &$S_{\rm 1D}$    \\
\hline
1	            & 0         & 1          & 0	      & 6 $\Gamma$  & 0    \\
2	            & 1	        & 0	         & 0          & 3 $\Gamma$  & 0.1  \\
3	            & 1/3	    & 1/3        & 1/3        & 3 $\Gamma$  & 0.2  \\
4               & 96/100    & 2/100      & 2/100      & 6 $\Gamma$  & 0.5  \\
5               & 1/2       & 1/2        & 0          & 3 $\Gamma$  & 1.2  \\
\hline
\end{tabular}
\caption{Experimental results showing the impact} of the power balance between the different sidebands on the effectiveness of 1D cooling, as measured by $S_{\rm 1D}$. $r_{\delta}$ is the fraction of ${\cal L}_{0}$ power used in sideband ${\cal S}_{\delta}$.
\label{tab:1Dsidebands}
\end{table}

The effectiveness of laser cooling depends on the intensity of the cooling light and on the overall detuning, $\Delta$, which is common to all three sidebands used to address the hyperfine components (see figure~\ref{fig:CoolingLevels}). The optimum detuning tends to increase as the intensity increases. However, the sidebands constrain the detuning that can be used. In particular, when $\Delta = 6\Gamma \approx 33$ MHz, ${\cal S}_{-192}$ is close to resonance with the $F=0,2$ components, which is  detrimental to cooling. This raises the question of how best to distribute the laser power between the three sidebands. Table~\ref{tab:1Dsidebands} shows the effectiveness of 1D laser cooling, given by $S_{\rm 1D}$, for five different sideband configurations. In all cases, $B=0.5$~G. In configurations 1 and 2, ${\cal L}_0$ has only one frequency component. For these configurations $S_{\rm 1D}$ is very small because the light is too far from resonance from some hyperfine components, resulting in a low photon scattering rate. In configuration 3, the power is distributed equally between the three sidebands, and we have chosen $\Delta=3\Gamma$. In this case, ${\cal S}_{-192}$ is detuned by $3\Gamma$ from the $F=1^+$ hyperfine state but by $-3\Gamma$ from the $F=0,2$ states. This configuration also results in poor cooling. In configuration 4, 96\% of the power is in ${\cal S}_0$, with the remainder divided equally between ${\cal S}_{-159}$ and ${\cal S}_{-192}$. This improves the cooling because all hyperfine components are addressed and ${\cal S}_{-192}$ is less detrimental to cooling because it has so little power. In configuration 5, we turn off ${\cal S}_{-192}$ altogether and divide the power equally between the other two sidebands. In this case, there is no longer any light red-detuned from the $F=0,2$ states, giving more freedom to optimize $\Delta$. The state $F=1^+$ can still be excited by ${\cal S}_{-159}$, which has high intensity. This is the best configuration we have found and is the one used in the rest of the paper.

\begin{figure}[tb]
\centering
\includegraphics[width=0.5\textwidth]{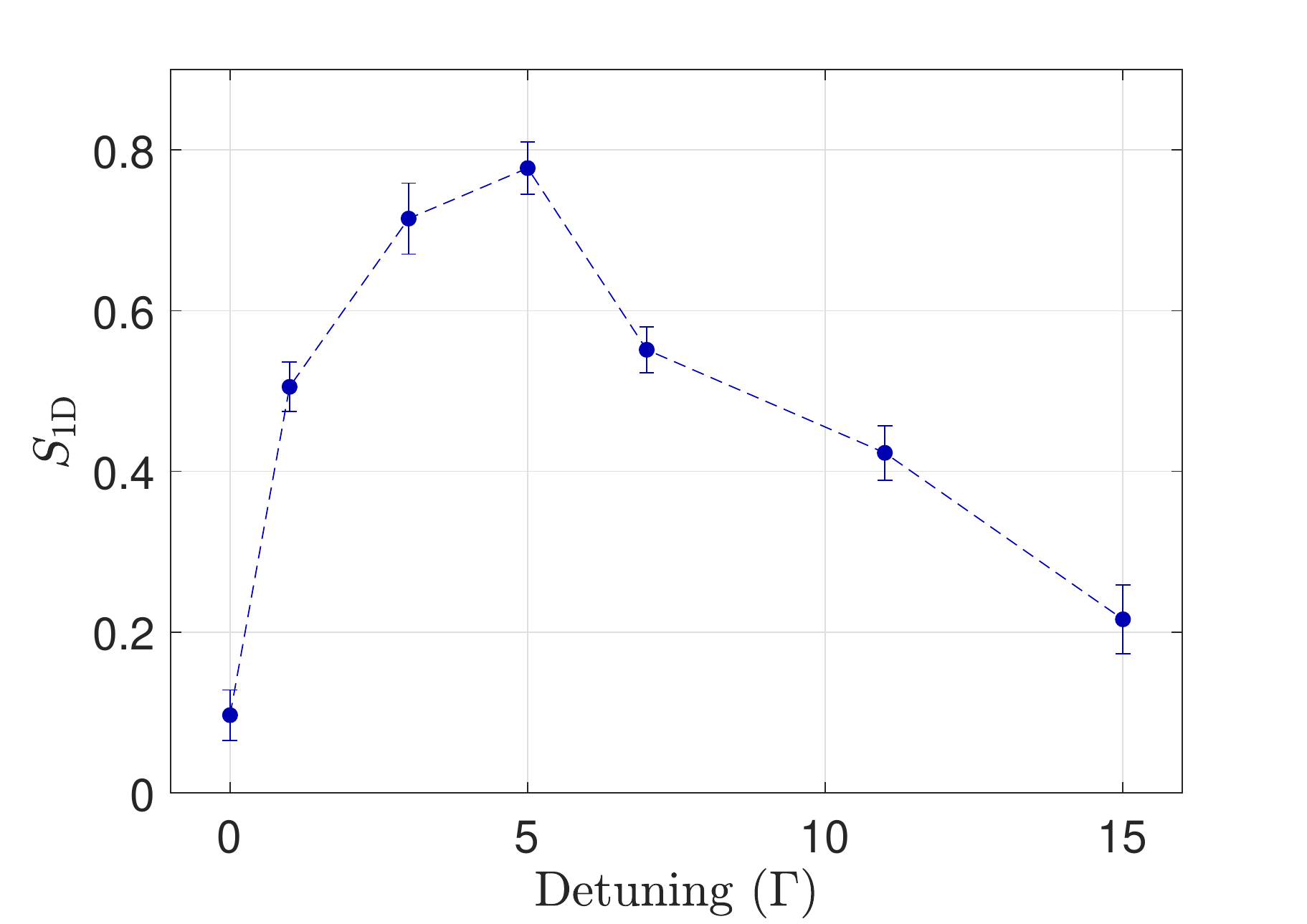}
\caption{$S_{\rm 1D}$ as a function of $\Delta$, with $B=0.5$~G and sideband configuration 5. $S_{\rm 1D}$ is obtained by fitting to data similar to figure \ref{fig:1Dprofile}(c), and error bars derive from the uncertainties in the fit parameters.}
\label{fig:1Ddetuning}
\end{figure}

Figure \ref{fig:1Ddetuning} shows  $S_{\rm 1D}$ as a function of $\Delta$, using sideband configuration 5, $B=0.5$~G, and a peak intensity per beam of $I_{\rm max} = 450 I_{\rm sat}$. The optimal detuning is about 5 $\Gamma$. Note that the $S_{\rm 1D}$ values for this set of data are smaller than typically obtained. The data were acquired at a different time, when the properties of the molecular beam may have been different. We have found that laser cooling results can be sensitive to the forward velocity of the beam and the temperature of the source.

\subsection{2D cooling}

\begin{figure}[tb]
\centering
\includegraphics[width=0.95\textwidth]{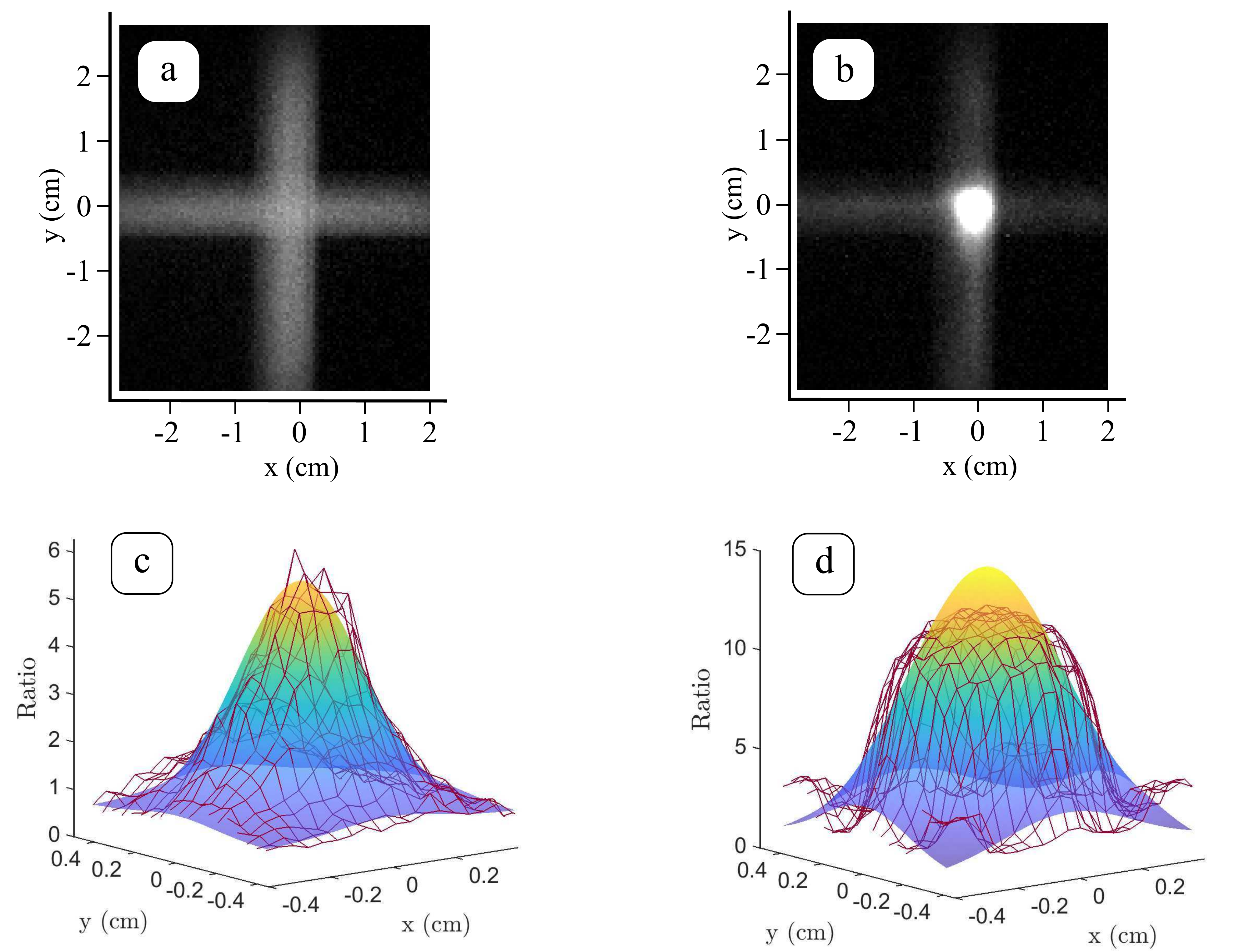}
\caption{Fluorescence images of molecules at CCD2. (a) Distribution without laser cooling. (b) Distribution with laser cooling in 2D. (c) Ratio of the two images. The mesh of lines show the data and the smooth surface is a 2D Gaussian fit. (d) Ratio found from the simulation results shown in fig \ref{fig:trajectorySimulation}(a). The mesh shows the simulation result and the smooth surface is a 2D Gaussian fit.}
\label{fig:2Dprofile}
\end{figure}

Figure \ref{fig:2Dprofile} shows typical data for laser cooling in two dimensions. Images (a) and (b) show the uncooled and cooled beam respectively. The bright spot at the centre of image (b) is an ultracold molecular beam. The density of molecules in this beam is increased due to the reduced divergence in both transverse directions. Figure~\ref{fig:2Dprofile}(c) shows the ratio of the two images. The mesh shows the data, and the smooth surface is a fit of the distribution to the 2D Gaussian model $g(x,y) = \frac{A}{2\pi}  e^{-(x-x_0)^2/(2\sigma_x^2)}e^{-(y-y_0)^2/(2\sigma_y^2)}+b$. As in the 1D case, about half of all molecules are lost from the cooling cycle. Despite this, the peak ratio is large; near the centre of the distribution laser cooling has increased the number of molecules by about a factor of 6. Following the procedure used in 1D, we define a signal $S_\mathrm{2D} =  A\sigma_x\sigma_y/L^2$, which is proportional to the number of ultracold molecules and is a useful measure when studying the effectiveness of 2D cooling as a function of the experimental parameters.
Figure~\ref{fig:2Dprofile}(d) shows the result of the simulations from figure \ref{fig:trajectorySimulation}(a). These simulations use the same parameters as used for the experiment, so we can make a direct comparison. The simulation data is treated in the same way as the experimental data, showing the ratio of the density distributions with and without cooling. The density with cooling is multiplied by 0.5 to reproduce the loss observed in the experiments. The simulation result, shown by the mesh in the figure, looks more like a top-hat profile than a Gaussian, but nevertheless we fit the same model $g(x,y)$.  The amplitude of the Gaussian fit is 2.8 times larger in the simulation than in the experiment, and the geometric mean width $(\sigma_x \sigma_y)^{1/2}$ is 1.2 times larger in the simulations. This suggests that the capture area is a little larger in the simulations than in reality, and that the effectiveness of cooling molecules that enter that area is also larger. These differences may be due to the approximations made in the trajectory simulations (see section \ref{sec:traj_sim}) or the assumptions made about the distribution exiting the source. Experimental imperfections such misalignments between the cooling beams could also be the cause. Nevertheless, given the complexity of the experiment and simulations, and the large parameter space, the comparison between simulation and experiment is reasonably good.

\subsubsection{Magnetic field and polarization}\label{sec:Bfield_Polar}

\begin{figure}[tb]
\centering
\includegraphics[width=0.5\textwidth]{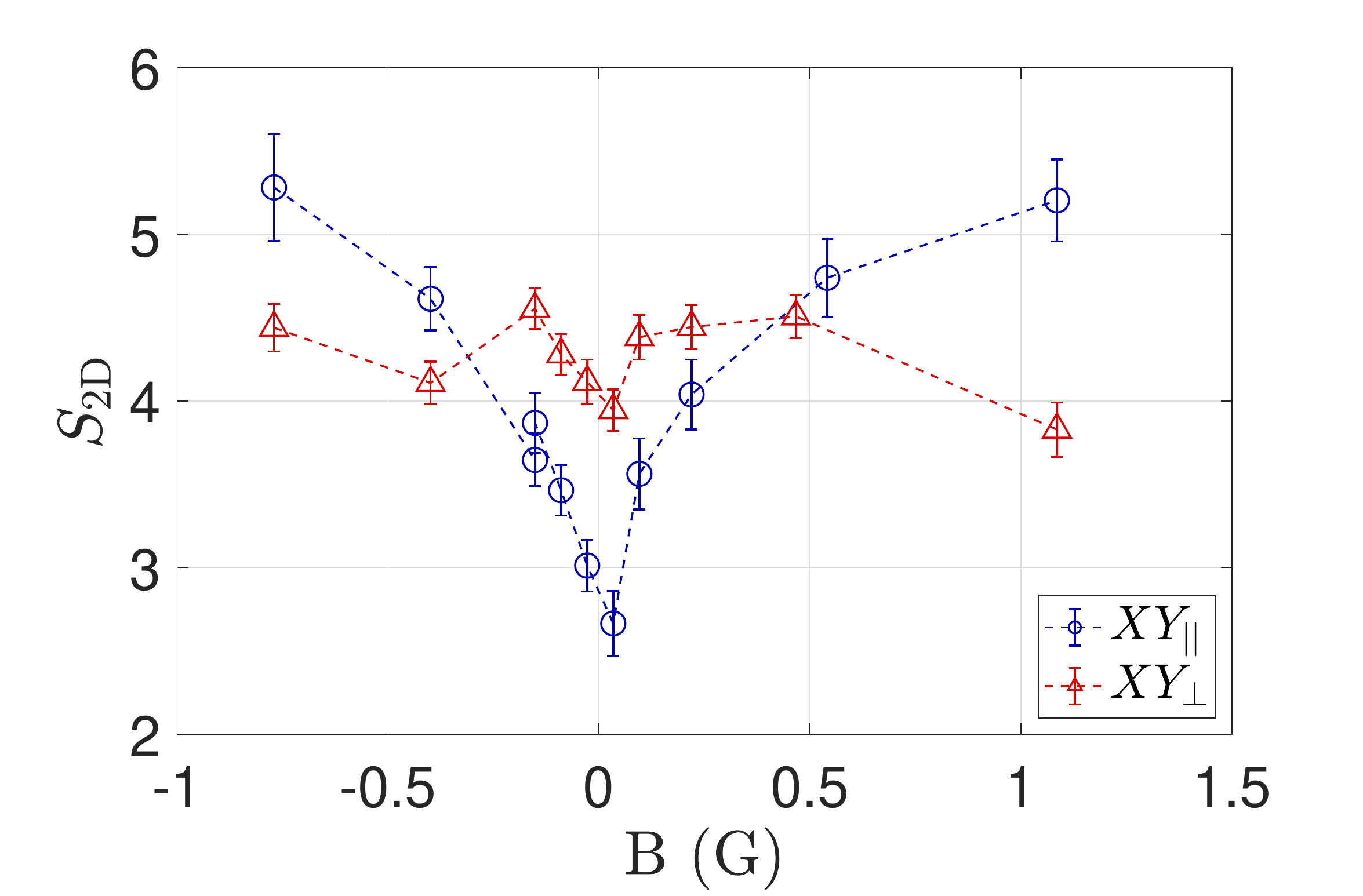}
\caption{$S_\mathrm{2D}$ as a function of $B$ with $\Delta= 6\Gamma$ and $I_{\rm max} = 450 I_{\rm sat}$. Blue circles: $XY_{||}$ configuration. Red triangles: $XY_{\perp}$ configuration.}
\label{fig:2DBfield}
\end{figure}

We have studied the effectiveness of the $XY_{||}$ and $XY_{\perp}$ configurations for a range of magnetic field values. Figure~\ref{fig:2DBfield} shows $S_\mathrm{2D}$ as a function of $B$ for these two polarization configurations. In the $XY_{||}$ case, the cooling mechanism is magnetically induced laser cooling. As expected, we see a sharp reduction in $S_\mathrm{2D}$ around $B=0$, but its value does not drop to zero as we might expect. The background magnetic field has been carefully cancelled, so this is most likely due to imperfect polarization of the laser beams. Indeed, we have seen that the polarization changes with position along the cooling region, due to stress-induced birefringence in the windows. Measurements going beyond the range shown in the plot indicate that the signal is roughly constant between 1 and 2.5~G. This is consistent with the intuitive picture given in section~\ref{sec:simulations}, which suggested an optimum $B$ of around 2~G. It is also consistent with the results presented in figure~\ref{fig:coolingForceMagneticField}(a), which shows that $v_0$ increases sharply between 0 and $1$~G, and then decrease slowly as $B$ is increased to $5$~G.

In the $XY_{\perp}$ configuration, we find that $S_\mathrm{2D}$ is independent of $B$ in the range $|B|< 1$~G. Notably, we see no significant change in cooling effectiveness around $B=0$. In this polarization configuration, there are both intensity and polarization gradients, so both polarization gradient cooling and (when $B\ne 0$) magnetically induced cooling can contribute. These results are consistent with the simulation results shown in figure~\ref{fig:coolingForceMagneticField}(b) which show no change to the acceleration curves over this range of $B$. We find that, provided $|B|>0.5$~G, the best results are obtained with the $XY_{||}$ configuration. This disagrees with the results shown in figure~\ref{fig:coolingForceMagneticField}, which suggest that the $XY_{\perp}$ configuration at low magnetic field produces a slightly larger critical velocity and maximum acceleration. This difference might be due to the polarization imperfections.
We restrict ourselves to the optimal $XY_{||}$ configuration for the rest of the paper.

\subsubsection{Detuning and laser intensity}
\begin{figure}[tb]
\centering
\includegraphics[width=0.5\textwidth]{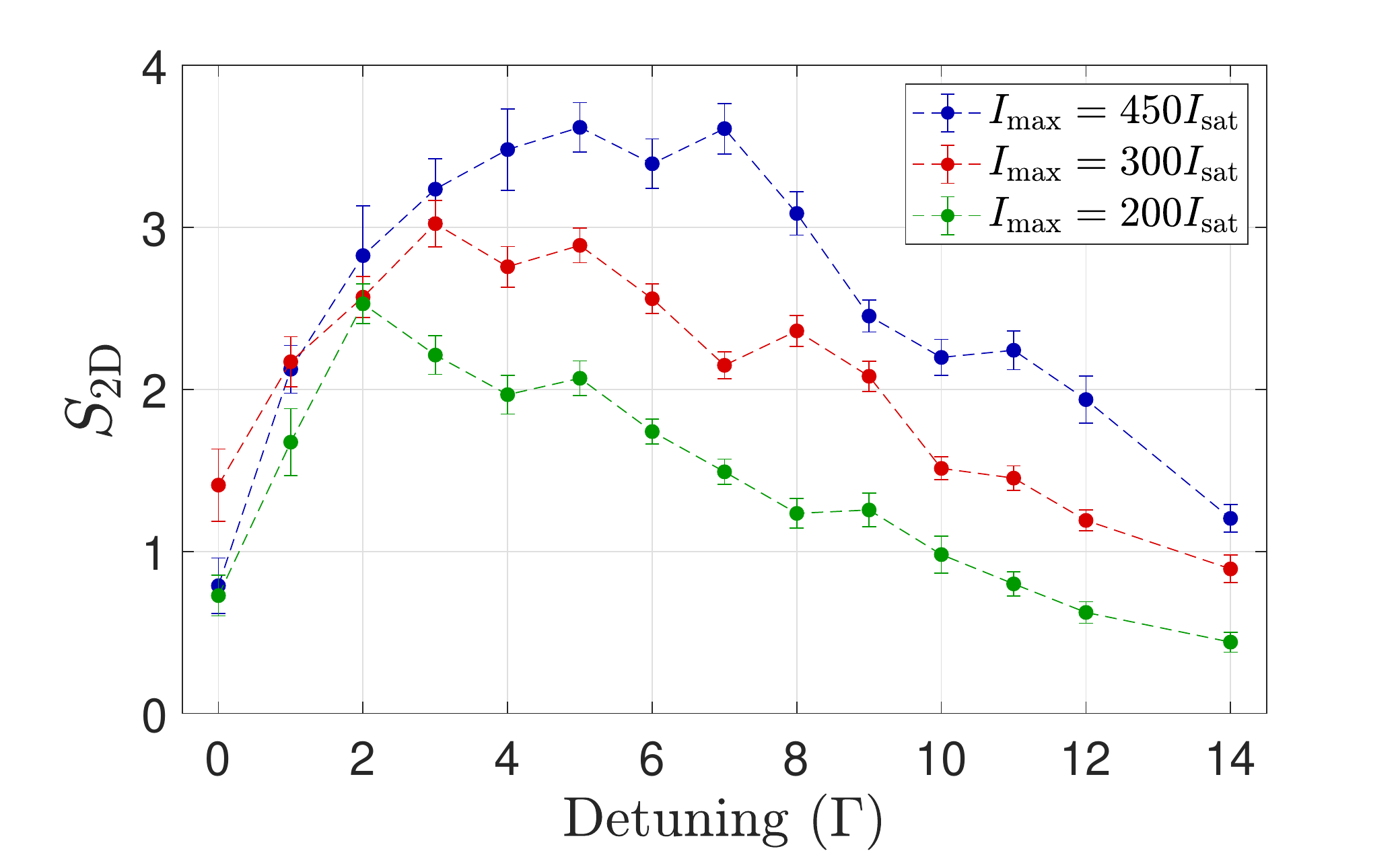}
\caption{$S_\mathrm{2D}$ as a function of $\Delta$ for three different values of the peak intensity per beam, $I_{\rm max}$. We have used $B=0.6$~G, the $XY_{||}$ polarization configuration, and sideband configuration 5 of table~\ref{tab:1Dsidebands}.}
\label{fig:2Ddetuning}
\end{figure}

Figure \ref{fig:2Ddetuning} shows $S_\mathrm{2D}$ as a function of $\Delta$ for three different values of $I_{\rm max}$. We see that the number of ultracold molecules increases with laser intensity, with no sign of saturating over this range of intensities. This is due to the increasing ac Stark shift producing higher potential hills and therefore more kinetic energy being lost per period, resulting in more effective cooling. The figure also shows that there is an optimal detuning, and that as $I_{\rm max}$ increases this optimal value also increases. This reflects the behaviour of the ac Stark shift, which is maximized at a higher detuning when the intensity is higher. Furthermore, increasing $\Delta$ lowers the excitation rate so that molecules have more time to climb the potential hills before being optically pumped to a dark state. Recalling that the peak intensity in the cooling region averaged along $z$ is about 0.24$I_{\rm max}$, we see that the continual increase of $S_{\rm 2D}$ with $I_{\rm max}$ is consistent with the simulation results shown in figure~\ref{fig:coolingForceIntensity}. At intermediate intensity the best results are obtained for $\Delta \approx 3\Gamma$, which is also consistent with the simulation results shown in figure~\ref{fig:coolingForceDetuning}. Finally, the observation that the optimum detuning shifts to higher values at higher intensities also agrees with our simulation results. 

For 2D cooling, we have investigated the additional sideband configurations described in section~\ref{sec:1D}. In all cases investigated, we find that ${\cal S}_{-192}$ is detrimental to cooling. In these measurements, we were able to independently scan the detunings of the two remaining sidebands, as well as scanning the power balance between them. While there are various choices of detunings and power balance which give similar results to configuration 5, none of these improve on that configuration.

\begin{figure}[tb]
\centering
\includegraphics[width=0.8\textwidth]{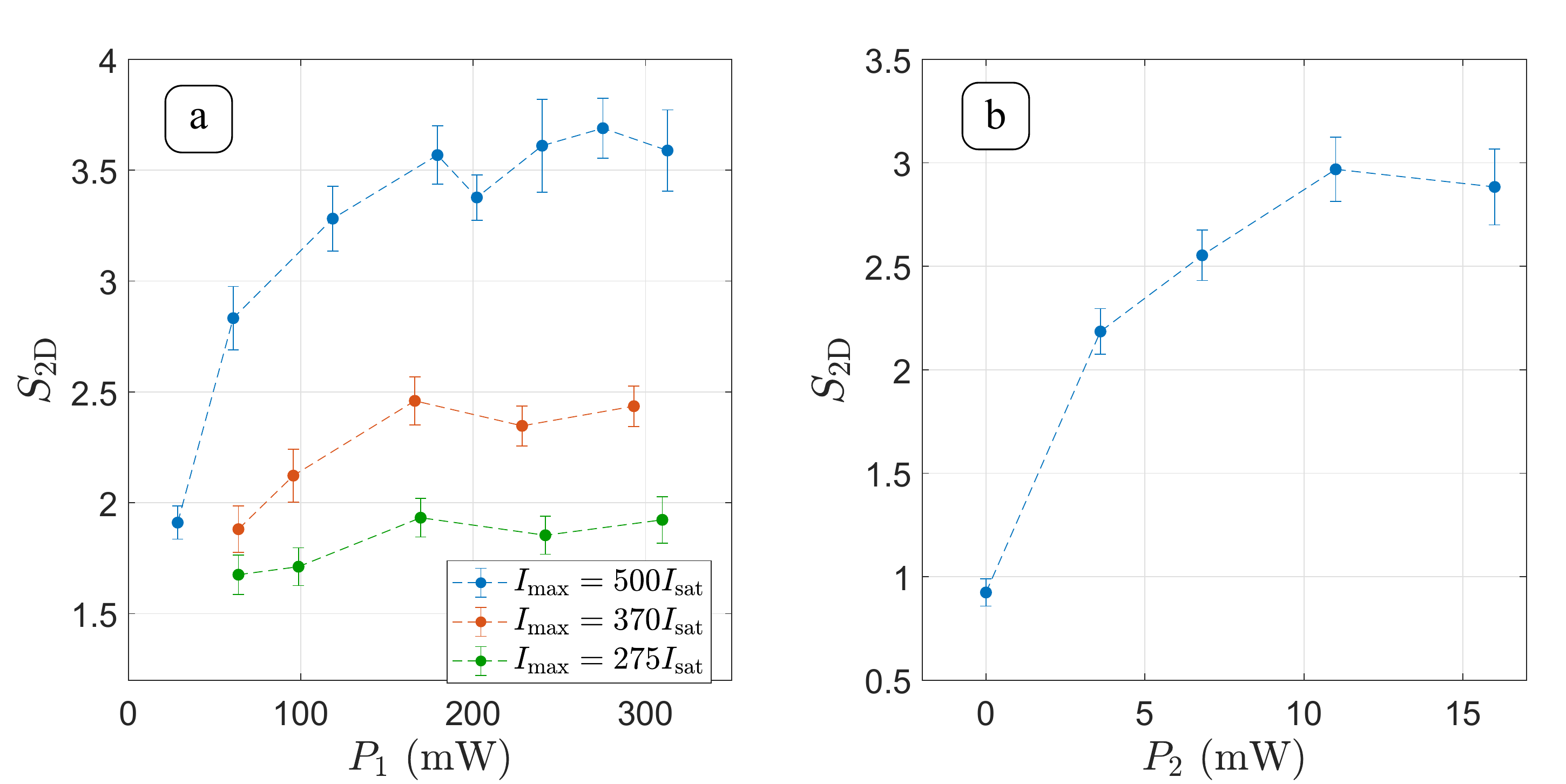}
\caption{(a) $S_\mathrm{2D}$ as a function of $P_1$, with $\Delta = 6\Gamma$ and for three different values of $I_{\rm max}$. $P_2$ is at its maximum value, (b) $S_\mathrm{2D}$ as a function of $P_2$, with $\Delta = 6\Gamma$ and $I_{\rm max}=500I_{\rm sat}$. $P_1$ is at its maximum value.}
\label{fig:v0v1v2powers}
\end{figure}

Figure \ref{fig:v0v1v2powers}(a) shows $S_\mathrm{2D}$ as a function of $P_1$ (the power of ${\cal L}_1$), for three different values of $I_{\rm max}$ and with $\Delta = 6 \Gamma$. The main conclusion we draw from these data is that the power available from ${\cal L}_1$ is sufficient to saturate the cooling, even at maximum $P_0$. Thus, this laser is not a limitation to the cooling. There is also some evidence that as $P_0$ increases, the value of $P_1$ needed to saturate the signal also increases. This is expected: increasing $P_0$ increases the rate at which molecules are pumped into $v=1$, so the excitation rate out of $v=1$ also has to be increased.  
Figure~\ref{fig:v0v1v2powers}(b) shows $S_\mathrm{2D}$ as a function of $P_2$ (the power of ${\cal L}_2$). Here, the signal saturates once $P_2$ exceeds 10~mW, and this laser is not a limitation to the cooling either.

\subsection{Temperature}\label{sec:temperature}

The temperature obtained in the experiments depends on several factors. If the molecules are cooled for long enough, they will reach an equilibrium temperature set by the ratio of the diffusion constant and the damping constant. The diffusion constant is determined by the photon scattering rate and the fluctuations of the dipole force. The damping constant is proportional to the gradient of the acceleration versus velocity curve near zero velocity (see figures \ref{fig:coolingForceMagneticField}, \ref{fig:coolingForceDetuning}, and \ref{fig:coolingForceIntensity}). The equilibrium temperature obtained from the simulations is typically around 10~$\mu$K. However, in our experiments there are many molecules captured by the molasses that are not cooled for a long enough time to reach the equilibrium temperature. This is easily seen from the simulation results by noting that the capture velocity is well above the critical velocity in all cases. Those molecules with speeds between $v_0$ and $v_{\rm capture}$, which we call the partly cooled molecules, can emerge from the laser cooling region with speeds 10--100 times higher than those that have reached equilibrium, so they have a strong influence on the measured temperature. The temperature will depend on the fraction of the beam made up by these partly-cooled molecules, how these molecules were distributed when they entered the cooling region, and how slow they are when they exit. It is difficult to predict how the temperature should depend on the control parameters in this complicated situation. However, we can expect there to be more partly cooled molecules when the capture velocity is high, since in this case there are more molecules that have not had enough time to reach the equilibrium temperature. When the capture velocity is high, there tends to be more molecules in the ultracold part of the distribution, so our main expectation is that the temperature will be higher when $S_{\rm 2D}$ is higher.

To estimate the transverse temperature of the molecular beam, we measure its density distribution at two positions using the two cameras labelled CCD1 and CCD2 in figure~\ref{fig:setup}. They are at distances of $z_1 = 0.9$~m and $z_2 = 1.5$~m from the end of the cooling region. respectively. The temperature is related to the rms widths, $\sigma_{1,2}$, of the Gaussian density distributions measured at these two detectors as
\begin{equation}\label{equ:temperature}
    T = \frac{m}{k_B} v_z^2 \frac{\sigma_2^2-\sigma_1^2}{z_2^2-z_1^2}.
\end{equation}
Here, $m$ is the mass of the molecule and $v_z$ is the average longitudinal velocity of the molecular beam, which is 200~m/s for the measurements presented here. At CCD2, we can determine the distribution in both the $x$ and $y$ directions, whereas at CCD1 only the distribution along $y$ is accessible. This means we can only measure temperatures in the $y$ direction. However, we note that the width of the distribution at CCD2 is typically the same in the two directions, and that the laser cooling region is identical in the two directions, so it is reasonable to assume that the temperature is the same in both transverse directions.

\begin{figure}[tb]
\centering
\includegraphics[width=1\textwidth]{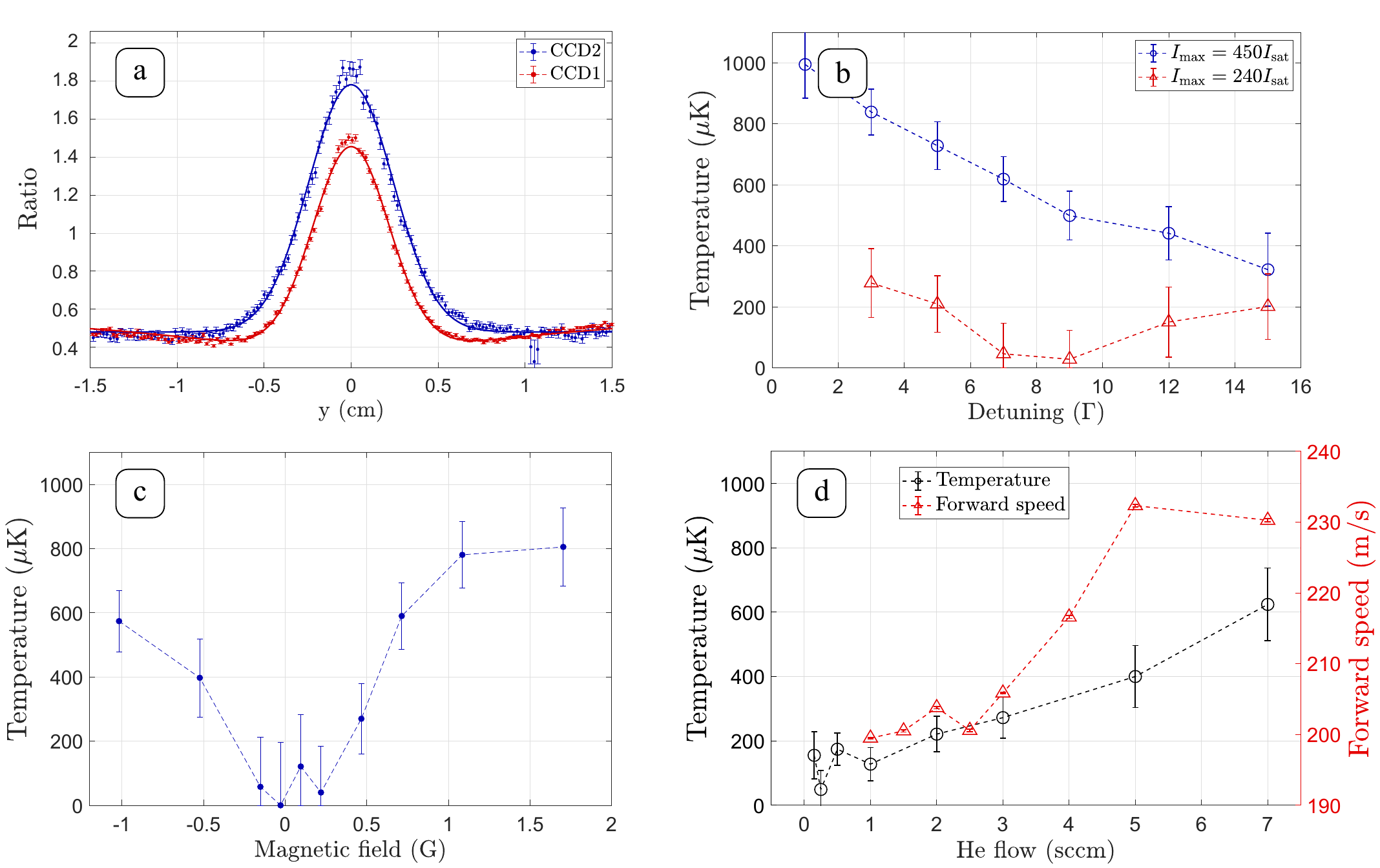}
\caption{(a) Profile of the cooled molecular beam in the $y$ direction, normalized to the uncooled distribution, at the positions $z_1 = 0.9$~m (CCD1) and $z_2 = 1.5$~m (CCD2). The solid lines are fits to the data (see text). (b) Temperature as a function of detuning $\Delta$, for two different values of $I_{\rm max}$ and with $B = 0.5$~G. (c) Temperature as a function of $B$ with $\Delta = 7 \Gamma$ and $I_{\rm max}=450 I_{\rm sat}$. (d) Temperature (black circles, left axis) and forward velocity (red triangles, right axis) as a function of the He buffer gas flow, with $\Delta = 7 \Gamma$ and $B = 0.55$~G.}
\label{fig:Temperatures}
\end{figure}

Figure \ref{fig:Temperatures}(a) shows example distributions measured at the two CCDs in a 2D cooling experiment. The data have been treated in the same way as for figures \ref{fig:1Dprofile} and \ref{fig:2Dprofile}. Note that the ratio has been averaged over the $x$ direction, so the peak ratio is smaller here than it is in figure \ref{fig:2Dprofile}. The uncooled molecules expand rapidly, so their number is smaller at CCD2 than at CCD1. By contrast, the number of ultracold molecules detected is the same at the two detectors, because this part of the beam is highly collimated. Consequently, after taking the ratio, the signal is larger at CCD2. At CCD1, we observe a small dip on each side of the central peak. These missing molecules are ones that have been cooled and have become part of the central peak. Because of this feature, we fit this distribution with a sum of two Gaussians with a common center. The first one has a positive amplitude and gives $\sigma_1$ (recall that the index 1 refers to CCD1), while the second one has a negative amplitude and a larger width. We do not observe these dips at CCD2 because the component with a negative amplitude has expanded too much to be noticeable. Therefore, we fit this distribution with a single Gaussian, which gives $\sigma_2$. For most of our data, the uncertainties in $\sigma_{1,2}$ are in the range 25--50~$\mu$m. This translates into a temperature uncertainty in the range 100--200~$\mu$K. As we will see, some of our measurements give temperatures consistent with zero within this uncertainty. 

Figure \ref{fig:Temperatures}(b) shows the influence of the detuning $\Delta$ on the transverse temperature for two different values of $I_{\rm max}$. At $240I_{\rm sat}$ the temperature decreases up to $\Delta \approx 9\Gamma$. Here, the results are limited by the temperature resolution of the measurement, but we obtain $T < 220$~$\mu$K with 95\% confidence. For larger detunings, the temperature increases again. At $450 I_{\rm sat}$ the temperature decreases from 1000 to 350~$\mu$K as $\Delta$ increases from 1 to $15\Gamma$, showing that the temperature increases with intensity and that the detuning needed to minimize the temperature is higher at higher intensity. While there is not a complete correlation, the trends in these results tend to follow the trends in the number of ultracold molecules -- when the number of molecules is large, the temperature also tends to be higher, consistent with the discussion above. This shows that there is a trade-off between maximizing the number of molecules and minimizing the temperature.

Figure \ref{fig:Temperatures}(c) shows how the temperature depends on the magnetic field. The trend observed is very similar to the one in figure~\ref{fig:2DBfield}. In particular, the temperature is minimized near $B=0$, which is also where the number of ultracold molecules is lowest. This is again consistent with the trend that conditions yielding more ultracold molecules also result in a higher temperature.

Figure \ref{fig:Temperatures}(d) shows how the temperature depends on the helium flow through the cryogenic buffer gas source (black circles). Over the range measured, we find that the temperature depends approximately linearly on the flow. A higher flow results in a faster beam, as shown in the figure (red triangles), so the molecules spend a shorter time in the laser cooling region. This is likely to be the reason for the increase in temperature, suggesting that a longer cooling region or a slower beam will help in reaching the lowest temperatures. The temperature may also be affected by the initial divergence of the beam, which tends to change as the flow changes, but we have not measured this. We typically obtain the most intense beam for flows around 2~sccm, and for this flow we measure a temperature of about 200~$\mu$K.

For optimized parameters, the temperature we achieve is below the resolution of our sensor. This is consistent with the simulations from figure \ref{fig:speedDistribution}, where the temperature reaches equilibrium in less than 1~ms. For parameters that optimize the number of molecules, the equilibrium temperature is not reached (figures \ref{fig:2Ddetuning} and \ref{fig:Temperatures}(b)). This could be remedied using a longer interaction region or a slower beam. The latter would also facilitate a longer eEDM measurement time.

\subsection{Beam brightness and implications for an eEDM measurement}

At present, the flux of molecules from the cryogenic source varies from week to week by up to a factor of 3. Our estimates in this section are based on a typical flux. When the laser cooling parameters are optimal, we obtain about $2 \times 10^{5}$ ultracold molecules per shot at CCD2, which is 1.5~m from the end of the cooling region. These molecules occupy an area about $5\times5$ mm$^2$, which is a convenient size for efficient detection. An ideal beam would contain a large number of molecules, have a small area, and a low divergence. Therefore, a good quantity to characterize the beam is its brightness (or radiance) which is defined as \cite{Lison1999}
${\cal R} = \frac{\dot{N}}{\Delta x^2\,\Delta\Omega}$ where $\dot{N}$ is the number of molecules per second passing through a small area $\Delta x^2$ into a solid angle $\Delta\Omega$. The solid angle is $\Delta\Omega = \pi(\Delta v_\perp/v_z)^2$, where $\Delta v_\perp = \sqrt{k_{\rm B} T/m}$ is the range of transverse velocities. The brightness of our ultracold beam is maximized at the lowest temperatures we measure ($T < 200$~$\mu$K), where we find ${\cal R} > 1\times10^{17}$ molecules/m$^2$/sr/s in the $N=1$ rotational state when the repetition rate is 10~Hz.  This lower limit on the brightness is about 300 times higher than without laser cooling.

Let us consider the precision that could be achieved by using this beam for an eEDM measurement. The statistical uncertainty of a measurement at the quantum projection noise limit is
\begin{equation}
    \sigma_{d_e} = \frac{1}{C}\frac{\hbar}{E_\mathrm{eff}\tau}\frac{1}{2\sqrt{N_\mathrm{mol}}}
    \label{eq:sigma_de}
\end{equation}
where $C$ is the contrast of the interferometer \cite{Ho2020}, $E_\mathrm{eff}$ is the effective electric field (see section \ref{sec:intro}), $\tau$ is the interrogation time and $N_\mathrm{mol}$ is the number of molecules detected.
At a temperature of $500$~$\mu$K or less, the molecules can travel for at least 50~ms without the transverse size of the beam increasing substantially. However, we take $\tau = 25$~ms, since it is limited at present by the beam speed and the length of the apparatus. Full use of the increased brightness could be achieved by slowing the molecular beam down using, e.g., longitudinal laser slowing, as discussed in the next section. With an applied electric field of 20~kV/cm we have $E_{\rm eff} = 17.5$~ GV/cm. We assume a repetition rate of 10~Hz, which is the same as used for the current experiments, a contrast of 0.9, a detection efficiency of 30\%, and a duty cycle of 0.5 to account for the dead time needed to reverse the electric field. With these values, equation (\ref{eq:sigma_de}) gives $\sigma_{d_e} = 5\times 10^{-30}$~e~cm after one day of measurement. A typical measurement of a hundred days brings the statistical uncertainty to $5 \times 10^{-31}$~e~cm. This estimate is 36 times more sensitive than the sensitivity realised using a supersonic source of YbF without laser cooling~\cite{Ho2020}.

\section{Summary and outlook}

In summary, we have applied 2D laser cooling to a beam of YbF molecules, producing a bright beam suitable for making an eEDM measurement with a statistical precision better than $10^{-30}$~e~cm. The cooling increases the brightness of the beam by a factor of about 300. We have explored how the number of ultracold molecules and their temperature depends on laser intensity, detuning and polarization configuration, and on the applied magnetic field. Detailed OBE simulations of the experiments help to interpret our experimental results and to build a good understanding of the cooling mechanisms at play and how they can be optimized. 

There are many ways to increase the number of ultracold molecules in the beam, improving eEDM sensitivity even further. As discussed in section \ref{sec:1D}, there appears to be a leak out of the cooling cycle that results in a loss of molecules. We are currently investigating the source of this loss using dispersed laser-induced fluorescence spectroscopy to identify leaks out of the cooling cycle. Removing the leak will increase the number of ultracold molecules in the beam by at least a factor of two.  We find that the time-averaged flux of molecules continues to increase as the repetition rate of the ablation laser is increased to its maximum value of 15~Hz. Recent work has shown that cryogenic buffer gas sources can be operated reliably up to a repetition rate of 55~Hz~\cite{Shaw2020}. Operation at such high repetition rates would be beneficial.  We are also investigating whether the flux can be improved using neon buffer gas. Neon can be pumped more effectively than helium, so the source can use higher flows of buffer gas which can lead to a large increase in the number of molecules~\cite{Hutzler2011, Hutzler2012}. We are also investigating whether the flux of molecules can be increased by exciting the Yb atoms in the source to the ${}^{3}P$ state. This has been shown to increase the flux of YbOH from a cryogenic buffer gas source by a factor of 10~\cite{Jadbabaie2020}. Simulations suggest that magnetic focussing of the molecules from the cryogenic source into the laser cooling region can increase the number brought to the ultracold regime by a factor of 20~\cite{Fitch2020b}. We are currently building the magnetic lens needed to do this. The temperature reached in this work is already low, but according to our simulations could be even lower. In particular, a small increase in the length of the cooling region, or decrease in beam speed, would help to ensure the molecules reach the equilibrium temperature, further increasing the brightness. Finally, we note that since the photon scattering rate is high and each molecule can scatter a large number of photons, a detection efficiency close to 1 is achievable, though we conservatively assumed only 30\% in our estimate above. 

At present, the interrogation time $\tau$ is limited by the speed of the beam, which was about $200$~m/s for most of the measurements presented here. This can be reduced by lowering the temperature from 6 to 4~K and optimizing the source design; mean speeds of 150~m/s are typical for this type of design. At this speed, $\tau = 25$~ms can be obtained using a 4~m long interaction region. This is a few times longer than molecular beam eEDM measurements built so far. The control and measurement of magnetic fields and gradients over this region will be especially challenging. To increase $\tau$ further, or reduce the length of the interaction region, the beam will need to be decelerated. Once the leak out of the cooling cycle has been closed, significant radiation pressure slowing becomes feasible~\cite{Barry2012, Truppe2017}. If the beam can be slowed to 30~m/s, the interrogation time can be increased to more than 100~ms and the sensitivity of the measurement can fully benefit from the increased brightness stated above. Alternatively, the molecules could be decelerated to rest and the eEDM measured using trapped molecules, as discussed in \cite{Fitch2020b}.

\ack
We thank Chris Ho for helpful discussions and feedback on this paper. We are grateful to Jon Dyne and David Pitman for their expert technical assistance. This research has received support from the Royal Society, the European Commission (grant 895187), and the Science and Technology Facilities Council (grant ST/S000011/1). The research was also partly supported by the John Templeton Foundation (grant 61104), the Gordon and Betty Moore Foundation (grant 8864), and the Alfred P. Sloan Foundation (grant G-2019-12505).

\section*{References}

\bibliographystyle{unsrtMax5}
\bibliography{references} %
\end{document}